\documentclass{statsoc}
\usepackage[left=7mm,right=15mm]{geometry} 
\usepackage{vmargin} 

\usepackage{amsmath,amssymb,hyperref}
\usepackage{natbib}
\usepackage{color}
\usepackage{pdfsync}
\usepackage{float}
\usepackage{graphicx}
\usepackage{subfigure}
\usepackage{xspace}
\usepackage[final]{pdfpages}

 \newtheorem{thm}{Theorem}
   \newtheorem{lem}[thm]{Proposition}

\newcommand{\x}{\mathbf{x}}
\newcommand{\h}{\mathbf{h}}
\newcommand{\myprod}{\prod_{\substack{i=1\\i\neq \h_t^n(s)}}^{N_x}}
\newcommand{\myprodone}{\prod_{\substack{i=1\\i\neq \h_t^n(1)}}^{N_x}}
\def\O{\mathcal{O}}
\newcommand{\SMCSQ}{SMC$^2$\xspace}
\newcommand{\LW}{L{\&}W\xspace}

\newcommand{\E}{\mathbb{E}} 
\newcommand{\iid}{\stackrel{iid}{\sim}}

\title[\SMCSQ]{\SMCSQ: an efficient algorithm for  sequential analysis of state-space models}

\author[Chopin et al.]{N. CHOPIN}
\address{CREST-ENSAE}
\email{nicolas.chopin@ensae.fr}
\author[Chopin et al.]{P.E. JACOB}
\address{CREST \& Universit\'e Paris Dauphine}
\email{pierre.jacob@ensae.fr}
\author[Chopin et al.]{O. PAPASPILIOPOULOS}
\address{Universitat Pompeu Fabra}
\email{omiros.papaspiliopoulos@upf.edu}

\begin{document}

\begin{abstract}
We consider the generic problem of performing sequential Bayesian inference
in a state-space model with observation process $y$, state
process $x$ and  fixed parameter $\theta$. 
An idealized approach would be to apply the iterated batch
importance sampling (IBIS) algorithm  of Chopin (2002). This
is a sequential Monte Carlo algorithm in the   $\theta$-dimension, 
that samples values of $\theta$, reweights iteratively these values
using the likelihood increments $p(y_t|y_{1:t-1},\theta)$, and rejuvenates
the $\theta$-particles through a resampling step and a MCMC update
step. In state-space models these likelihood increments are intractable
in most cases, but they may be unbiasedly estimated by a particle
filter in the $x$-dimension, for any fixed $\theta$. This
motivates the \SMCSQ algorithm proposed in this article: a sequential Monte
Carlo algorithm, defined in the $\theta$-dimension, which propagates and
resamples many particle filters in the $x$-dimension. The filters in
the $x$-dimension are an example of the random weight particle filter
as in Fearnhead et al. (2010). On the other hand, the particle
Markov chain Monte Carlo (PMCMC) framework
developed in Andrieu et al. (2010) allows us to design appropriate MCMC rejuvenation
steps. Thus, the $\theta$-particles target the
correct posterior distribution at each iteration $t$, despite the
intractability of the likelihood increments.  
We explore the applicability of our algorithm in both sequential and
non-sequential applications and  consider various degrees of freedom,
as for example increasing dynamically the 
number of $x$-particles.  We contrast our approach to various
competing methods, both conceptually and empirically through a
detailed simulation study, included here and
in a supplement, and based on particularly challenging examples. 


\end{abstract}

\section{Introduction}

\subsection{Objectives}

We consider a generic state-space model, with parameters $\theta\in\Theta$,
prior $p(\theta)$, latent Markov process $(x_{t})$,
$p(x_{1}|\theta)=\mu_{\theta}(x_{1})$,
\[
p(x_{t+1}|x_{1:t},\theta)=p(x_{t+1}|x_{t},\theta)=f_{\theta}(x_{t+1}|x_{t}),
\quad t\geq1,\]
and observed process\[
p(y_{t}|y_{1:t-1},x_{1:t-1},\theta)=p(y_{t}|x_{t},\theta)=g_{\theta}(y_{t}|x_{t}
),\quad t\geq1.\]
For an overview of such models with references to a wide range of
applications in Engineering, Economics, Natural Sciences, and other fields,
see e.g. \citet{DouFreiGor}, \citet{Kun:SSHMM} or \citet{CapMouRyd}.

We are interested in  the
recursive exploration of the sequence of  posterior distributions
\begin{equation}
\label{eq:recursion}
\pi_{0}(\theta)=p(\theta),\quad\pi_{t}(\theta,x_{1:t})=p(\theta,x_{1:t}|y_{1:t})
,\quad
t\geq1\,,
\end{equation}
as well as computing the model evidence $p(y_{1:t})$ for model
composition.  
Such a sequential analysis of state-space models under parameter uncertainty 
is of interest in many settings; a simple example is out-of-sample prediction, 
and related goodness-of-fit diagnostics based on prediction residuals, 
which are popular for instance in Econometrics; see e.g. Section 4.3 of 
\citet{kim1998stochastic} or \citet{Koop2007}.
Furthermore, we shall see that 
recursive exploration up to time $t=T$ may be computationally advantageous
even in batch estimation scenarios, where a fixed observation record $y_{1:T}$ is available. 

\subsection{State of the art}

Sequential Monte Carlo (SMC) methods are considered the state of the art for
tackling this kind of problems. 
Their appeal lies in the efficient
re-use of samples across different times $t$, compared for example
with MCMC methods which would typically have to be re-run for each
time horizon.  Additionally, convergence properties (with respect to the number of 
simulations) under mild assumptions are now well understood; see e.g. \citet{DelGui}, 
\citet{crisan2002survey}, \citet{Chopin:CLT}, \citet{OudRub}, \citet{douc:moulines:2008}. 
See also \citet{DelDouJas:SMC} for a recent overview of SMC methods. 

SMC methods are particularly (and rather unarguably) effective for
exploring the simpler sequence of posteriors, $\pi_{t}(x_t
|\theta)=p(x_{t}|y_{1:t},\theta)$; compared to the general case the static
parameters
are treated as known and interest is focused on $x_t$ as opposed to
the whole path $x_{0:t}$. This is typically called the filtering
problem. The
 corresponding algorithms are
known  as \emph{particle filters} (PFs); they are described in
Section \ref{sub:PartFiltering-x} in some detail. These algorithms
evolve, weight and resample a population of $N_x$ number of
particles, $x_t^{1:N_x}$, so that at each time $t$ they are a properly
weighted sample from  $\pi_{t}(x_t
|\theta)$. Recall that a particle system is called properly weighted if the
weights associated with each sample are unbiased estimates of the
Radon-Nikodym derivative between the target and the proposal
distribution; see for example Section 1 of \citet{high} and references
therein. A by-product of the PF output is an unbiased estimator of the likelihood
increments and the marginal likelihood
\begin{equation}
p(y_{1:t}|\theta)=p(y_{1}|\theta)\prod_{s=2}^{t}p(y_{s}|y_{1:s-1},\theta),
\quad1\leq
t\leq T.\label{eq:trueZt}
\end{equation}
the variance of which increases linearly over time \citep{CerDelGuy2011nonasymptotic}.

Complementary to this setting is the iterated batch importance
sampling (IBIS) algorithm of \citet{Chopin:IBIS} for the recursive exploration of the
sequence
of parameter posterior distributions,
$\pi_t(\theta)$; the algorithm is outlined in Section
\ref{sub:An-idealized-algorithm}. This is also an SMC algorithm which updates a
population of $N_\theta$ particles, $\theta^{1:N_\theta}$, so that at each time
$t$ they are a properly
weighted sample from  $\pi_t(\theta)$. The algorithm includes
occasional MCMC steps for rejuvenating the current population of
$\theta$-particles to prevent the number of distinct
$\theta$-particles from decreasing over time. Implementation of the
algorithm requires the  likelihood increments
$p(y_{t}|y_{1:t-1},\theta)$ to be computable. This constrains the
application of IBIS in state-space
models since computing the increments 
involves  integrating out  the latent
states. Notable exceptions are linear Gaussian state-space models and
models where $x_t$ takes values in a finite set. In such cases
a Kalman filter and a Baum filter respectively can be associated to each
$\theta$-particle to evaluate efficiently the likelihood
increments; see e.g. \citet{Chopin2007}. 

On the other hand, sequential inference for both parameters and latent
states for a generic state-space model is a much harder problem, which, although
very important in
applications, is still rather unresolved; see for example
\citet{DoucPoyiaSin}, \citet{PMCMC}, \citet{Doucetetal2009OverviewParameterEstimation} 
for recent discussions. The batch estimation problem
of exploring $\pi_T(\theta,x_{0:T})$ is  a non-trivial MCMC
problem on its own right, especially for large $T$. This is due
to both high dependence between parameters and the latent process, which
affects Gibbs sampling strategies \citep{pap:rob:sk}, and the
difficulty in designing efficient simulation schemes for sampling from
$\pi_T(x_{0:T} |\theta)$.  To address these problems \citet{PMCMC} developed a
general theory of 
particle Markov chain Monte Carlo (PMCMC) algorithms, which are MCMC
algorithms that use a particle filter of size $N_x$ as 
a proposal mechanism. Superficially, it appears that the algorithm
replaces the intractable \eqref{eq:trueZt} by the unbiased estimator
provided by the PF within an MCMC algorithm that samples from
$\pi_T(\theta)$. However, 
 \citet{PMCMC} show that (a) as $N_x$ grows,
the PMCMC algorithm behaves more and more like the theoretical MCMC
algorithm which targets the intractable $\pi_T(\theta)$; and (b) for any fixed
value of $N_x$, the PMCMC algorithm 
admits $\pi_T(\theta,x_{0:T})$ as a stationary distribution. The
exactness (in terms of not perturbing the stationary distribution)
follows from demonstrating that the PMCMC is an ordinary MCMC
algorithm (with specific proposal distributions) on an expanded model
which includes the PF as auxiliary variables; when  $N_x=1$ this
augmentation collapses to the more familiar scheme of imputing the
latent states.

\subsection{Proposed algorithm}

\SMCSQ is a generic black box tool for performing sequential analysis
of state-space models, which can be seen as  a natural extension of both IBIS and
PMCMC. To each of the $N_\theta$ $\theta-$particles $\theta^{m}$, we
attach a PF which propagates $N_x$ $x-$particles; due to the nested
filters we call it the \SMCSQ algorithm.  Unlike the
implementation of IBIS which carries an exact filter, in this case the
PFs only produce 
unbiased estimates of the marginal likelihood.  This  ensures 
 that the $\theta$-particles are properly
weighted for $\pi_t(\theta)$, in the spirit of the \emph{random
  weight PF} of e.g. \citet{high}. The connection with the auxiliary
representation underlying PMCMC is pivotal for designing the MCMC
rejuvenation steps, which are crucial for the success of IBIS. 
We obtain a sequential auxiliary Markov representation, 
and use it to formally demonstrate that our algorithm explores the sequence defined in
\eqref{eq:recursion}. The case $N_x=\infty$ corresponds to an
(unrealisable) IBIS algorithm, whereas $N_x=1$ to
an importance sampling scheme, the variance of which typically 
grows 
polynomially with $t$ \citep{Chopin:CLT}. 

\SMCSQ is a sequential but not an on-line algorithm. The
computational load increases with iterations due to the associated
cost of the MCMC steps. Nevertheless, these steps typically occur at a
decreasing rate (see Section 3.8 for details).  
The only on-line generic algorithm for sequential analysis
of state-space models we are aware of is the self-organizing
particle filter (SOPF) of 
\cite{Kitagawa:Self}: this is PF applied to  the extended state $\tilde{x}_t=(x_t,\theta)$, 
which never updates the $\theta$-component of particles, and typically diverges quickly
over time \citep[e.g.][]{Doucetetal2009OverviewParameterEstimation}; see also \cite{LiuWest} 
for a modification of SOPF which we discuss later. Thus, a genuinely on-line
analysis, which would provide constant Monte Carlo error at a constant CPU cost, with respect to all
the components of $(x_{1:t},\theta)$ may well be an unattainable goal. 
This is unfortunate, but hardly surprising, given that 
the target $\pi_t$ is of increasing dimension. 
For certain models with a specific structure (e.g the existence of sufficient statistics),  an on-line
algorithm may be obtained by extending SOPF so as to include MCMC updates of the $\theta$-component,
see \cite{GilksBerzu},  \cite{Fearnhead:Sufficient}, \cite{Storvik}, and also the more recent work of 
\cite{carvalho2010particlelearning},  but numerical evidence
seems to indicate these algorithms degenerate as well, albeit possibly at a slower rate; see e.g. 
\citet{Doucetetal2009OverviewParameterEstimation}. 
On the other hand, \SMCSQ is a generic approach which does not require such a specific structure. 
 
Even in batch estimation scenarios \SMCSQ may offer several
advantages over PMCMC, in the same way that
SMC approaches may be advantageous over MCMC methods
\citep{Neal:AIS,Chopin:IBIS,Robert:PMC,DelDouJas:SMC,jasra2007population}.  
Under certain conditions (which relate to the
asymptotic normality of the maximizer of \eqref{eq:trueZt}) \SMCSQ 
has the same complexity as PMCMC.
Nevertheless, it calibrates
automatically its tuning parameters, as for example $N_x$ and the
proposal distributions for $\theta$. (Note adaptive versions 
of PMCMC, see e.g. \cite{silva2009particle} and \cite{PetersHosackHayes} exist however.) 
Then, the first iterations of the \SMCSQ
algorithm make it possible
to quickly discard uninteresting parts of the sampling space, using
only a small number of observations. Finally, 
the \SMCSQ algorithm provides an estimate
of the evidence (marginal likelihood) of the model $p(y_{1:T})$  as a direct by-product. 

We demonstrate the potential of the \SMCSQ on two classes of problems
which involve multidimensional state processes and several parameters:
volatility prediction for financial assets using L\'evy driven
stochastic volatility models, and likelihood assessment  of athletic
records using time-varying extreme value distributions. A supplement
to this article (available on the third author's web-page) 
contains further numerical investigations with the
\SMCSQ and competing methods on more standard examples.  

Finally, it has been pointed to us that \cite{FulopLi} have developed independently
and concurrently an algorithm similar to \SMCSQ. Distinctive features 
of our paper are the generality of the proposed approach, so that it 
may be used more or less automatically on complex examples (e.g. 
setting $N_x$ dynamically), and the formal results that establish the validity of the \SMCSQ
algorithm, and its complexity.

\subsection{Plan, notations}

The paper is organised as follows. Section \ref{sec:PF_IBIS} recalls the two basic
ingredients of \SMCSQ: the PF and the IBIS.  Section \ref{sec:smc2}
introduces the \SMCSQ algorithm, provides its formal justification,
discusses its complexity and the latitude in its implementation.  Section
\ref{sec:numerics} carries out a detailed simulation study which investigates the
performance of \SMCSQ on particularly challenging models. Section \ref{sec:conclusion}
concludes.  

As above, we shall use extensively the concise colon notation for sets of random variables, e.g. 
$x_t^{1:N_x}$ is a set of $N_x$ random variables $x_t^{n}$, $n=1,\ldots,N_x$,
$x_{1:t}^{1:N_x}$ is the union of the sets  $x_{s}^{1:N_x}$, $s=1,\ldots,t$, and so on. In the same
vein, ${1:N_x}$ stands for the set $\{1,\ldots,N_x\}$.  Particle 
(resp. time) indices are always in superscript (resp. subscript). The letter $p$ refers
to probability densities defined by the model, e.g. $p(\theta)$, $p(y_{1:t}|\theta)$, while
$\pi_t$ refers to the probability density targeted at time $t$ by the algorithm, or the corresponding
marginal density with respect to its arguments. 

\section{Preliminaries} \label{sec:PF_IBIS}

\subsection{Particle filters (PFs)}
\label{sub:PartFiltering-x}

We describe a
particle filter that approximates recursively the sequence of filtering
densities $\pi_t(x_t|\theta)=p(x_{t}|y_{1:t},\theta)$, for a fixed parameter
value
$\theta$. The formalism is chosen with view to integrating this
algorithm into \SMCSQ.  
We first give a pseudo-code version, and then we detail the notations.
Any operation involving the superscript  $n$ 
must be understood as performed for $n\in 1:N_x$, where $N_x$ is
the total number of particles. 
\medskip\hrule\medskip
Step 1: At iteration $t=1$, 
\begin{description}
\item [{(a)}] Sample $x_{1}^{n}\sim q_{1,\theta}(\cdot)$. 
\item [{(b)}] Compute and normalise weights\[
w_{1,\theta}(x_{1}^{n})=\frac{\mu_{\theta}(x_{1}^{n})g_{\theta}(y_{1}|x_{1}^{n})
}{q_{1,\theta}(x_{1}^{n})},\quad
W_{1,\theta}^{n}=\frac{w_{1,\theta}(x_{1}^{n})}{\sum_{i=1}^{N_x}w_{1,\theta}(x_{1}
^{i})}.\]

\end{description}
Step 2: At iteration $t=2:T$,
\begin{description}
\item [{(a)}] Sample the index $a_{t-1}^{n} \sim
\mathcal{M}(W_{t-1,\theta}^{1:N_x})$ of the ancestor of particle
$n$. 
\item [{(b)}] Sample $x_{t}^{n}\sim q_{t,\theta}(\cdot|x_{t-1}^{a_{t-1}^{n}})$. 
\item [{(c)}] Compute and normalise weights\[
w_{t,\theta}(x_{t-1}^{a_{t-1}^{n}},x_{t}^{n})=\frac{f_{\theta}(x_{t}^{n}|x_{t-1}
^{a_{t-1}^{n}})g_{\theta}(y_{t}|x_{t}^{n})}{q_{t,\theta}(x_{t}^{n}|x_{t-1}^{a_{
t-1}^{n}})},\quad
W_{t,\theta}^{n}=\frac{w_{t,\theta}(x_{t-1}^{a_{t-1}^{n}},x_{t}^{n})}{\sum_{i=1}^{N_x}w_{
t,\theta}(x_{t-1}^{a_{t-1}^{i}},x_{t}^{i})}.\]
\end{description}
\medskip\hrule\medskip

In this algorithm, $\mathcal{M}(W_{t-1,\theta}^{1:N_x})$ stands for
the multinomial distribution which assigns probability $W_{t-1,\theta}^{n}$
to outcome $n\in 1:N_x$,
and $\left(q_{t,\theta}\right)_{t\in 1:T}$ 
stands for a sequence of conditional proposal distributions which
depend on $\theta$. A standard, albeit sub-optimal, choice is the
prior, $q_{1,\theta}(x_{1})=\mu_{\theta}(x_{1})$,
$q_{t,\theta}(x_{t}|x_{t-1})=f_{\theta}(x_{t}|x_{t-1})$
for $t\geq2$, which leads to the simplification
$w_{t,\theta}(x_{t-1}^{a_{t-1}^{n}},x_{t}^{n})=g_{\theta}(y_{t}|x_{t}^n)$.
We note in passing that Step (a) is equivalent to multinomial resampling
\citep[e.g.][]{Gordon}. Other, more efficient schemes exist
\citep{LiuChen,Kitagawa:Self,CarClifFearn},
but are not discussed in the paper for the sake of
simplicity. 

At iteration $t$, the following quantity 
\begin{equation*} 
\frac{1}{N_x}\sum_{n=1}^{N_x}w_{t,\theta}(x_{t-1}^{a^n_{t-1}}, x_t^n)
\end{equation*}
is an unbiased estimator of $p(y_{t}|y_{1:t-1},\theta)$. More generally, 
it is a key feature of PFs that   
\begin{equation} \label{eq:Zhatt}
\hat{Z}_{t}(\theta,x_{1:t}^{1:N_x},a_{1:t-1}^{1:N_x})=
\left(\frac{1}{N_x}\right)^{t}
\left\{\sum_{n=1}^{N_x}w_{1,\theta}(x_{1}^{n})\right\} \prod_{s=2}^{t}\left\{
\sum_{n=1}^{N_x}w_{s,\theta}(x_{s-1}^{a_{s-1}^{n}},x_{s}^{n})\right\} 
\end{equation}
is also an unbiased estimator of $p(y_{1:t}|\theta)$; this is not a
straightforward result, see  Proposition 7.4.1 in \citet{delMoral:book}.
We denote by $\psi_{1,\theta}(x_{1}^{1:N_x})$, for $t=1$, and $
\psi_{t,\theta}(x_{1:t}^{1:N_x},a_{1:t-1}^{1:N_x})$
 for $t\geq2$, the joint probability density of all the random variables
generated during the course of the algorithm up to iteration $t$. 
Thus, the expectation of the random variable
$\hat{Z}_{t}(\theta,x_{1:t}^{1:N_x},a_{1:t-1}^{1:N_x})$
with respect to $\psi_{t,\theta}$ is exactly $p(y_{1:t}|\theta)$.

\subsection{Iterated batch importance sampling (IBIS) 
\label{sub:An-idealized-algorithm}}

The  IBIS approach of
\citet{Chopin:IBIS} is an SMC algorithm for exploring a sequence of
parameter posterior distributions $\pi_t(\theta) = p(\theta |
y_{1:t})$.  
All the operations involving the particle index $m$ must be understood
as operations performed for all $m\in 1: N_\theta$, where
$N_\theta$ is the total number of $\theta$-particles.  
\medskip\hrule\medskip
Sample $\theta^{m}$ from $p(\theta)$ and set $\omega^m\leftarrow1$.
Then, at time $t=1:T$
\begin{description}
\item [{(a)}] Compute the incremental weights and their weighted average\[
u_{t}(\theta^{m})=p(y_{t}|y_{1:t-1},\theta^{m}),\quad
L_t=\frac{1}{\sum_{m=1}^{N_\theta}\omega^m}\times\sum_{m=1}^{N_\theta}\omega^m
 u_t(\theta^m),\]
with the convention  $p(y_{1}|y_{1:0},\theta)= p(y_{1}|\theta)$
for $t=1$.
\item [{(b)}] Update the importance weights,\begin{equation}
\omega^{m}\leftarrow\omega^{m}u_{t}(\theta^{m}).\label{eq:isideal}\end{equation}

\item [{(c)}] If some degeneracy criterion is fulfilled, sample
$\tilde{\theta}^{m}$
independently from the mixture distribution \[
\frac{1}{\sum_{m=1}^{N_\theta}\omega^{m}}\sum_{m=1}^{N_\theta}\omega^{m}K_{t}
\left(\theta^{m},\cdot
\right).\]
 Finally, replace the current weighted particle system,
by the set of new, unweighted particles: \[
(\theta^{m},\omega^{m})\leftarrow(\tilde{\theta}^{m},1).\]
\end{description}
\medskip\hrule\medskip

\citet{Chopin:CLT} shows that  \[
\frac{\sum_{m=1}^{N_\theta}\omega^{m}\varphi(\theta^m)}{\sum_{m=1}^{N_\theta}
\omega^{m}}\]
is a consistent and asymptotically (as $N_\theta \to \infty$) normal
estimator of the expectations
$$\E\left[\varphi(\theta)|y_{1:t}\right]=\int\varphi(\theta)p(\theta|y_{1:t})\,
d\theta,$$ for all appropriately integrable $\varphi$. In addition, each $L_t$,  computed
in Step (a), is a consistent
and asymptotically normal estimator of the likelihood 
$p(y_t|y_{1:t-1})$. 

 Step (c) is usually decomposed
into a resampling and a mutation step. In the above algorithm the
former is done with the multinomial distribution, 
 where particles are selected with probability proportional
to $\omega^{m}$. As mentioned in Section
\ref{sub:PartFiltering-x} other resampling schemes may be used
instead.  The move step is achieved through a Markov kernel $K_{t}$ which leaves
$p(\theta|y_{1:t})$
invariant. In our examples $K_t$ will be a Metropolis-Hastings kernel. A
significant advantage of IBIS is that the population of
$\theta$-particles can be used to learn features of the
target distribution, e.g by computing
\[
\widehat{\Sigma}=\frac{1}{\sum_{m=1}^{N_\theta}\omega^m }\sum_{m=1}^{N_\theta}
\omega^m \left(\theta^{m}-\hat{\mu}\right)\left(\theta^{m}-\hat{\mu}\right)^{T}
,\quad\hat{\mu}=\frac{1}{\sum_{m=1}^{N_\theta}\omega^m }\sum_{m=1}^{N_\theta}
\omega^m \theta^{m}.\]
 New particles can be proposed according to a
Gaussian random walk $\tilde{\theta}^m|\theta^m\sim
N(\theta^m,c\widehat{\Sigma})$, 
where $c$ is a tuning constant for
achieving optimal scaling of the Metropolis-Hastings
algorithm, or independently  $\tilde{\theta}^m \sim
N(\hat{\mu},\widehat{\Sigma})$ as  suggested in
\citet{Chopin:IBIS}. 
A standard degeneracy criterion is
$\mathrm{ESS}<\gamma N_\theta$, for $\gamma \in (0,1)$, where  ESS stands for ``effective
sample size'' and is computed as
\begin{equation}
  \label{eq:ess}
  \mathrm{ESS}=\frac{\left(\sum_{m=1}^{N_\theta}\omega^{m}\right)^{2}}{\sum_{m=1}^{N_\theta
}\left(\omega^{m}\right)^{2}}.
\end{equation}
Theory and practical guidance on the use of
this criterion are provided in Sections \ref{sub:complex} and
\ref{sec:numerics} respectively.

In the context of state-space models IBIS is a
theoretical algorithm since the likelihood increments $p(y_{t}|y_{1:t-1},\theta)$
(used both in Step 2, and implicitly in the MCMC kernel) are typically
intractable. Nevertheless, coupling IBIS with PFs yields a working algorithm as
we show in the following section. 

\section{Sequential parameter and state estimation: the \SMCSQ algorithm}
\label{sec:smc2}

\SMCSQ is a natural amalgamation of IBIS and PF. We first provide the
algorithm, we then demonstrate its validity and we close the section
by considering various possibilities in its implementation. 
Again, all the operations involving the index $m$ must be understood
as operations performed for all $m\in 1:N_\theta$. 
\medskip\hrule\medskip
Sample $\theta^{m}$ from $p(\theta)$ and set $\omega^{m}\leftarrow1$.
Then, at time $t=1,\ldots,T$,
\begin{description}
\item [{(a)}] For each particle $\theta^{m}$, perform iteration $t$ of
the PF described in Section \ref{sub:PartFiltering-x}:
If $t=1$, sample independently $x_{1}^{1:N_x,m}$ from
 $\psi_{1,\theta^{m}}$, and compute 
\[
\hat{p}(y_{1}|\theta^{m})=\frac{1}{N_x}\sum_{n=1}^{N_x}w_{1,\theta}(x_{1}^{n,m});\]
If $t>1$, sample $ \left(x_{t}^{1:N_x,m}, a_{t-1}^{1:N_x,m}\right)$
from $\psi_{t,\theta^{m}}$ conditional on
$\left(x_{1:t-1}^{1:N_x,m},a_{1:t-2}^{1:N_x,m}\right)$,
and compute 
\[
\hat{p}(y_{t}|y_{1:t-1},\theta^{m})=\frac{1}{N_x}\sum_{n=1}^{N_x}w_{t,\theta}(x_{t-1}^{
a_{t-1}^{n,m},m},x_{t}^{n,m}).\]

\item [{(b)}] Update the importance weights,
\begin{equation}
\omega^{m}\leftarrow\omega^{m}\hat{p}(y_{t}|y_{1:t-1},\theta^{m}).\label{eq:ishat}
\end{equation}

\item [{(c)}] If some degeneracy criterion is fulfilled, sample
$\left(\tilde{\theta}^{m},\tilde{x}_{1:t}^{1:N_x,m},\tilde{a}_{1:t-1}^{1:N_x}
\right)$
independently from the mixture distribution \[
\frac{1}{\sum_{m=1}^{N_\theta}\omega^{m}}\sum_{m=1}^{N_\theta}\omega^{m}K_{t}
\left\{
  \left(\theta^{m},x_{1:t}^{1:N_x,m},a_{1:t-1}^{1:N_x,m}\right), \cdot
\right \} \]
where $K_{t}$ is a PMCMC kernel described in Section \ref{sub:PMCMCstep}.
Finally, replace the current weighted particle system by the set
of new unweighted particles: \[
(\theta^{m},x_{1:t}^{1:N_x,m},a_{1:t-1}^{1:N_x,m},\omega^{m})\leftarrow(\tilde{
\theta}^{m},\tilde{x}_{1:t}^{1:N_x,m},\tilde{a}_{1:t-1}^{1:N_x,m},1).\]
\end{description}
\medskip\hrule\medskip
The degeneracy criterion in Step (c) will typically be the same as for
IBIS, i.e., when the ESS
drops below a threshold, where the ESS is computed as in
\eqref{eq:ess} and the 
$\omega^m$'s are now obtained in \eqref{eq:ishat}. We study the stability and the
computational cost of the algorithm when applying this criterion in Section
\ref{sub:complex}. 

\subsection{Formal justification of \SMCSQ}

A proper formalisation of the successive importance sampling steps
performed by the \SMCSQ algorithm requires extending the sampling
space, in order to include all the random variables generated by the
algorithm. 

At time $t=1$, the algorithm generates variables $\theta^{m}$ from
the prior $p(\theta)$, and for each $\theta^{m}$, the algorithm
generates vectors $x_{1}^{1:N_x,m}$ of particles, from
$\psi_{1,\theta^{m}}(x_{1}^{1:N_x})$.
Thus, the sampling space is $\Theta\times\mathcal{X}^{N_x}$, and the
actual ``particles'' of the algorithm are $N_\theta$ independent and
identically distributed copies of the random variable $(\theta,x_{1}^{1:N_x})$,
with density:\[
p(\theta)\psi_{1,\theta}(x_{1}^{1:N_x})=p(\theta)\prod_{n=1}^{N_x}q_{1,\theta}
(x_{1}^{n}).\]
Then, these particles are assigned importance weights corresponding
to the incremental weight function
$\hat{Z}_{1}(\theta,x_{1}^{1:N_x})=N_x^{-1}\sum_{n=1}^{N_x}w_{1,\theta}(x_{1}^{n
})$. This means that, at iteration 1, the target distribution of the algorithm
should be defined as:
$$
\pi_{1}(\theta,x_{1}^{1:N_x}) =
p(\theta)\psi_{1,\theta}(x_{1}^{1:N_x}) \times
\frac{\hat{Z}_{1}(\theta,x_{1}^{1:N_x})}{p(y_1)},
$$
where the normalising constant $p(y_1)$ is easily deduced from the property that 
$\hat{Z}_{1}(\theta,x_{1}^{1:N_x})$ is an unbiased estimator of $p(y_1|\theta)$. 
To understand the properties of $\pi_{1}$, simple manipulations suffice.
Substituting $w_{1,\theta}(x_{1}^{n})$, $\psi_{1,\theta}(x_{1}^{1:N_x})$ and 
$\hat{Z}_{1}(\theta,x_{1}^{1:N_x})$ with their respective expressions,
\begin{eqnarray*}
\pi_{1}(\theta,x_{1}^{1:N_x}) & = &
\frac{p(\theta)}{p(y_1)}\prod_{i=1}^{N_x}q_{1,\theta}(x_{1}^{i})\left\{ 
\frac{1}{N_x}\sum_{n=1}^{N_x}\frac{\mu_{\theta}(x_{1}^{n})g_{\theta}(y_{1}|x_{1}^{n})}
{q_{1,\theta}(x_{1}^{n})}\right\}
\\ 
 & = &
\frac{1}{N_x}\sum_{n=1}^{N_x}\frac{p(\theta)}{p(y_1)}\mu_{\theta}(x_{1}^{n})g_{\theta}(y_{1}|x_{1}^{n}
)\left\{ \prod_{i=1,i\neq n}^{N_x}q_{1,\theta}(x_{1}^{i})\right\}
\end{eqnarray*}
 and noting that, for the triplet $(\theta,x_{1},y_{1})$ of random
variables, \[
p(\theta)\mu_{\theta}(x_{1})g_{\theta}(y_{1}|x_{1})=p(\theta,x_{1},y_{1})=p(y_{1
})p(\theta|y_{1})p(x_{1}|y_{1},\theta)\]
one finally gets that: 
\begin{eqnarray*}
\pi_{1}(\theta,x_{1}^{1:N_x}) & = &
\frac{p(\theta|y_{1})}{N_x}\sum_{n=1}^{N_x}p(x_{1}^{n}|y_{1},\theta)\left\{
\prod_{i=1,i\neq n}^{N_x}q_{1,\theta}(x_{1}^{i})\right\} .
\end{eqnarray*}
The following two properties of $\pi_{1}$ are easily deduced from
this expression. First, the marginal distribution of $\theta$ is
$p(\theta|y_{1})$. Thus, at iteration 1 the algorithm
is properly weighted  for any
 $N_x$.  
Second, conditional on $\theta$, $\pi_{1}$ assigns to the vector
$x_{1}^{1:N_x}$ a mixture distribution which  with probability $1/N_x$,
gives to particle $n$ the filtering distribution $p(x_{1}|y_{1},\theta)$,
and to all the remaining particles the proposal distribution
$q_{1, \theta}$. The notation reflects these properties by  denoting the
target 
 distribution of \SMCSQ by $\pi_1$, since 
 it admits the distributions defined in \eqref{eq:recursion} as
 marginals.

By a simple induction, one sees that the target density $\pi_{t}$
at iteration $t\geq2$ should be defined as:
\begin{equation}\label{eq:def_pit}
\pi_{t}(\theta,x_{1:t}^{1:N_x},a_{1:t-1}^{1:N_x}) = 
p(\theta)\psi_{t,\theta}(x_{1:t}^{1:N_x},a_{1:t-1}^{1:N_x})
\times
\frac{\hat{Z}_{t}(\theta,x_{1:t}^{1:N_x},
a_{1:t-1}^{1:N_x})}{p(y_{1:t})}
\end{equation}
where $\hat{Z}_{t}(\theta,x_{1:t}^{1:N_x},
a_{1:t-1}^{1:N_x})$ was defined in \eqref{eq:Zhatt}, 
that is, it should be proportional to the sampling density of
all random variables generated so far, times the product of the successive
incremental weights. Again, the normalising constant $p(y_{1:t})$ 
in \eqref{eq:def_pit} is easily deduced from the fact
that $\hat{Z}_{t}(\theta,x_{1:t}^{1:N_x},
a_{1:t-1}^{1:N_x})$ is an unbiased estimator of $p(y_{1:t}|\theta)$. 
The following Proposition gives an alternative expression
for $\pi_{t}$.

\begin{lem}
\label{lem:pit}The probability density $\pi_{t}$ may be written
as:
\begin{eqnarray}
\lefteqn{\pi_{t}(\theta,x_{1:t}^{1:N_x},a_{1:t-1}^{1:N_x})=
p(\theta|y_{1:t})\times}\label{eq:prop_pit}\\
 &  & \frac{1}{N_x} \sum_{n=1}^{N_x}\frac{p(\x_{1:t}^{n}|\theta,y_{1:t})}{N_x^{t-1}}
\left\{ \myprodone q_{1,\theta}(x_{1}^i)\right\} \left\{ \prod_{s=2}^t
\myprod W_{s-1,\theta}^{a_{s-1}^{i}}q_{s,\theta}(x_s^i|x_{s-1}^{a_{s-1}^{i}})\right\} 
\nonumber 
\end{eqnarray}
where $\x_{1:t}^{n}$ and $\mathbf{h}_{t}^{n}$ are deterministic
functions of $x_{1:t}^{1:N_x}$ and $a_{1:t-1}^{1:N_x}$ defined as follows:
$\mathbf{h}_{t}^{n}=\left(\mathbf{h}_{t}^{n}(1),\ldots,\mathbf{h}_{t}^{n}
(t)\right)$
denote the index history of $x_{t}^{n}$, that is, $\mathbf{h}_{t}^{n}(t)=n$, and
$\mathbf{h}_{t}^{n}(s)=a_s^{\mathbf{h}_{t}^n(s+1)}$, recursively, for $s=t-1,\ldots,1$, 
and
$\x_{1:t}^{n}=\left(\x_{1:t}^{n}(1),\ldots,\x_{1:t}^{n}(t)\right)$
denote the state trajectory of particle $x_{t}^{n}$, i.e.
$\x_{1:t}^{n}(s)=x_{s}^{\h_{t}^{n}(s)}$,
for $s=1,\ldots,t$.
\end{lem}

A proof is given in Appendix A. We use a bold notation 
to stress out that 
the quantities  $\x_{1:t}^{n}$ and $\mathbf{h}_{t}^{n}$ are quite different
from particle arrays such as e.g. $x_{1:t}^{1:N_x}$: 
$\x_{1:t}^{n}$ and $\mathbf{h}_{t}^{n}$ provide the complete genealogy 
of the particle with label $n$ at time $t$, while  $x_{1:t}^{1:N_x}$ 
simply concatenates the successive particle arrays $x_{t}^{1:N_x}$, and contains no such 
genealogical information. 

It follows immediately from expression \eqref{eq:prop_pit} that the marginal distribution
of $\pi_{t}$ with respect to $\theta$ is  $p(\theta|y_{1:t})$.  Conditional on
$\theta$
the remaining random variables, $x_{1:t}^{1:N_x}$ and $a_{1:t-1}^{1:N_x}$,
have a mixture distribution, according to which, with probability
$1/N_x$ the  state trajectory $\x_{1:t}^{n}$ is generated according to
$p(x_{1:t}|\theta,y_{1:t})$, the ancestor variables corresponding to this 
trajectory, $a_s^{\h_t^n(s)}$ are uniformly distributed within ${1:N_x}$, 
and all the other random variables are
generated from the particle filter  proposal distribution, $\psi_{t,\theta}$. 
Therefore, Proposition \ref{lem:pit} establishes a sequence of auxiliary distributions
$\pi_t$ on increasing dimensions, whose marginals include the
posterior distributions of interest defined in
\eqref{eq:recursion}. The \SMCSQ algorithm targets this sequence using SMC
techniques.

\subsection{The MCMC rejuvenation step}
\label{sub:PMCMCstep}

To formally describe  this step performed at some iteration $t$, we must work,
as in
the previous section, on the extended set of variables 
$(\theta,x_{1:t}^{1:N_x},a_{1:t-1}^{1:N_x})$. The 
algorithm is described below; if the proposed move is accepted, the
set of variables is replaced by the proposed one, otherwise it is left
unchanged. The algorithm is based on some proposal kernel $T(\theta,d\tilde{\theta})$
in the $\theta-$dimension, which admits probability density $T(\theta,\tilde{\theta})$. 
(The proposal kernel for $\theta$, $T(\theta,\cdot)$, may be chosen as described
in Section \ref{sub:An-idealized-algorithm}.)
\medskip\hrule\medskip
\begin{description}
\item [{(a)}] Sample $\tilde{\theta}$ from proposal kernel, $\tilde{\theta}\sim
T(\theta,d\tilde{\theta})$. 
\item [{(b)}] Run a new PF for $\tilde{\theta}$:  sample
independently  $(\tilde{x}_{1:t}^{1:N_x},\tilde{a}_{1:t-1}^{1:N_x})$ from
$\psi_{t,\tilde{\theta}}$, and compute
$\hat{Z}_{t}(\tilde{\theta},\tilde{x}_{1:t}^{1:N_x},\tilde{a}_{1:t-1}^{1:N_x})$.
\item [{(c)}] Accept the move with probability 
\[
1\wedge\frac{p(\tilde{\theta})\hat{Z}_{t}(\tilde{\theta},\tilde{x}_{1:t}^{1:N_x}
,\tilde{a}_{1:t-1}^{1:N_x})T(\tilde{\theta},\theta)}{p(\theta)\hat{Z}_{t}(\theta
,x_{1:t}^{1:N_x},a_{1:t-1}^{1:N_x})T(\theta,\tilde{\theta})}\,.\] 
\end{description}
\medskip\hrule\medskip

It directly follows from  \eqref{eq:def_pit} that this algorithm
defines a standard Hastings-Metropolis kernel
with proposal distribution
\[
q_{\theta}(\tilde{\theta},\tilde{x}_{1:t}^{1:N_x},\tilde{a}_{1:t}^{1:N_x}
)=T(\theta,\tilde{\theta})\psi_{t,\tilde{\theta}}(\tilde{x}_{1:t}^{1:N_x},\tilde
{a}_{1:t}^{1:N_x})
\]
 and admits as invariant distribution the extended distribution
 $\pi_{t}(\theta,x_{1:t}^{1:N_x},a_{1:t-1}^{1:N_x})$. 
In the broad PMCMC framework, this scheme corresponds to
the so-called particle Metropolis-Hastings algorithm
\citep[see][]{PMCMC}. It is worth pointing out an interesting
digression from the PMCMC framework. The Markov mutation kernel has to be
invariant with respect to $\pi_t$, but it does not necessarily need to
produce an ergodic Markov chain, since consistency of Monte Carlo
estimates is achieved by averaging across many particles and not within a path of a
single particle. Hence, we can also attempt lower dimensional updates, e.g
using a Hastings-within-Gibbs algorithm. The advantage of such moves
is that they might lead to higher acceptance rates for the same step
size in the $\theta$-dimension. However, we do not pursue this point
further in this article. 

\subsection{PMCMC's invariant distribution, state inference}\label{sub:invariant}

From \eqref{eq:prop_pit}, one may rewrite $\pi_t$ as the marginal
distribution of $(\theta,x_{1:t}^{1:N_x},a_{1:t-1}^{1:N_x})$ with respect 
to an extended distribution that would include a uniformly distributed particle index $n^\star\in 1:N_x$: 
\begin{eqnarray}
\lefteqn{\pi_t^\star(n^\star,\theta,x_{1:t}^{1:N_x},a_{1:t-1}^{1:N_x})
=\frac{p(\theta|y_{1:t})}{N_x^t}\times} \nonumber\\
 &  & p(\x_{1:t}^{n^\star}|\theta,y_{1:t})\left\{ 
\prod_{\stackrel{i=1}{i\neq \h_t^{n^\star}(1)}}^{N_x} q_{1,\theta}(x_{1}^i)\right\} \left\{ \prod_{s=2}^t
\prod_{\stackrel{i=1}{i\neq \h_t^{n^\star}(s)}}^{N_x}
W_{s-1,\theta}^{a_{s-1}^{i}}q_{s,\theta}(x_s^i|x_{s-1}^{a_{s-1}^{i}})\right\} 
\label{eq:Andrieuetal_pit}.
\end{eqnarray}

\citet{PMCMC} formalise PMCMC algorithms as MCMC algorithms that leaves $\pi_t^\star$ invariant, 
whereas in the previous section we justified our PMCMC update as a MCMC step leaving $\pi_t$ invariant. 
This distinction is a mere technicality in the PMCMC context, but it becomes 
important in the sequential context. \SMCSQ is best understood as an algorithm
targetting the sequence $(\pi_t)$: defining importance sampling steps between
successive versions of $\pi_t^\star$ seems cumbersome, as the interpretation 
of $n^\star$ at time $t$ does not carry over to iteration $t+1$. This distinction
also relates to the concept of Rao-Blackwellised (marginalised) particle filters \citep{DouGodAnd}:
since $\pi_t$ is a marginal distribution with respect to $\pi_t^\star$, targetting $\pi_t$
rather than $\pi_t^\star$ leads to more efficient (in terms of Monte Carlo variance) SMC
algorithms. 

The interplay between $\pi_t$ and  $\pi_t^\star$  is exploited below and in the
following sections in order to fully realize the implementation
potential of \SMCSQ.
As a first example, direct inspection of  \eqref{eq:Andrieuetal_pit} reveals that 
the conditional distribution of $n^\star$, given $\theta$, $x_{1:t}^{1:N_x}$ and 
$a_{1:t-1}^{1:N_x}$, is  $\mathcal{M}(W_{t,\theta}^{1:N_x})$, the multinomial distribution that assigns probability 
$W_{t,\theta}^n$ to outcome $n$, $n\in 1:N_x$. Therefore, weighted samples
from $p(\theta,x_{1:t}|y_{1:t})$ may be obtained at iteration $t$
as follows: 
\medskip\hrule\medskip 
\begin{description}
\item [{(a)}] For $m=1,\ldots,N_\theta$, draw index $n^{\star}(m)$ from $\mathcal{M}(W_{t,\theta^m}^{1:N_x})$.
\item [{(b)}] Return the weighted sample \[
(\omega^{m},\theta^{m},\x_{1:t}^{n^{\star}(m),m})_{m\in 1:N_\theta}\]
where $\x_{1:t}^{n,m}$ was defined in Proposition
\ref{lem:pit}.
\end{description}
\medskip\hrule\medskip
This \emph{temporarily extended} particle system can be used in the standard way to
make inferences about $x_t$ (filtering), $y_{t+1}$ (prediction) or even $x_{1:t}$ (smoothing), 
under parameter uncertainty. Smoothing requires to 
store all the state variables $x_{1:t}^{1:N_x,1:N_\theta}$, which is expensive, 
but filtering and prediction may be performed while storing only the most recent 
state variables,  $x_{t}^{1:N_x,1:N_\theta}$. We discuss more thoroughy the memory cost 
of \SMCSQ, and explain how smoothing may still be carried out at certain
times, without storing the complete trajectories, in Section \ref{sub:complex}. 

The phrase \emph{temporarily extended} in the previous paragraph refers to our discussion on the difference 
between $\pi_t$ and $\pi_t^\star$.
By extending the particles with a $n^\star$ component, one temporarily change the target
distribution, from $\pi_t$ to $\pi_t^\star$. To propagate to time $t+1$, 
one must revert back to $\pi_t$, by simply marginaling out the particle index
$n^\star$. We note however that, before reverting to $\pi_t$, one has the liberty to apply MCMC updates 
with respect to $\pi_t^\star$. For instance, one may
update the $\theta-$component of each particle according to the full
conditional distribution of $\theta$ with respect to to $\pi_t^\star$, 
that is, $p(\theta|\x_{1:t}^{n^\star},y_{1:t})$. Of course, 
this possibility is interesting mostly for those models such 
that $p(\theta|\x_{1:t}^{n^\star},y_{1:t})$ is tractable. 
And, again, this operation may be performed only if all the state variables are available
in memory.

\subsection{Reusing all the $x-$particles}

The previous section describes an algorithm for obtaining a particle sample 
$(\omega^{m},\theta^{m},\x_{1:t}^{n^{\star}(m),m})_{m\in 1:N_\theta}$ that targets 
$p(\theta,x_{1:t}|y_{1:t})$. One may use this sample to compute, 
for any test function $h(\theta,x_{1:t})$, an estimator
of the expectation of $h$ with respect to the target $p(\theta,x_{1:t}|y_{1:t})$: 
\begin{equation*}
\frac{1}{\sum_{m=1}^{N_\theta} \omega^m } \sum_{m=1}^{N_\theta} \omega^m h(\theta^m,\x_{1:t}^{n^\star(m),m}).
\end{equation*}

\noindent As in \citet[][Section 4.6]{PMCMC}, we may deduce from this expression a Rao-Blackwellised
estimator, by marginalising out $n^\star$, and re-using all the $x$-particles: 
\begin{equation*}
\frac{1}{\sum_{m=1}^{N_\theta} \omega^m }\sum_{m=1}^{N_\theta} \omega^m 
\left\{\sum_{n=1}^{N_x} W_{t,\theta^m}^n  h(\theta^m,\x_{1:t}^{n,m})
\right\}.
\end{equation*}

\noindent The variance reduction obtained by this Rao-Blackwellisation scheme  should depend on the variability of  
$h(\theta^m,\x_{1:t}^{n,m})$  with respect to $n$.
For a fixed $m$, the components $\x_{1:t}^{n,m}(s)$ of the trajectories $\x_{1:t}^{n,m}$ are diverse 
when $s$ is close to $t$, and degenerate when $s$ is small. Thus, this Rao-Blackwellisation
scheme should be more efficient when $h$ depends mostly on recent state
values, e.g. $h(\theta,x_{1:t})=h(x_t)$, and less efficient when $h$ depends mostly on early state
values, e.g. $h(\theta,x_{1:t})=h(x_1)$.

\subsection{Evidence}

The evidence  of the data obtained
up to time $t$ may be decomposed using the chain rule: 
\[
p(y_{1:t})=\prod_{s=1}^{t}p(y_{s}|y_{1:s-1}).\]
The IBIS algorithm delivers the weighted averages $L_s$, for each 
$s=1,\ldots,t$, which are
Monte Carlo estimates of the corresponding factors in the
product; see Section \ref{sub:An-idealized-algorithm}. 
Thus, it provides an estimate of the evidence by multiplying
these terms. This can also
be achieved via the \SMCSQ algorithm in a similar manner:
\[
\hat{L}_t=\frac{1}{\sum_{m=1}^{N_\theta}\omega^{m}}\sum_{m=1}^{N_\theta}\omega^{
m}\hat{p}(y_{t}|y_{1:t-1},\theta^{m})\]
where $\hat{p}(y_{t}|y_{1:t-1},\theta^{m})$ is given in the definition
of the algorithm. It is therefore possible to
estimate the evidence of the model, at each iteration
$t$, at practically no extra cost.

\subsection{Automatic calibration of $N_x$}
\label{sec:increase}

The plain vanilla \SMCSQ algorithm assumes that $N_x$ 
stays  constant during the complete run. This poses two practical difficulties.
First, choosing
a moderate value of $N_x$ that leads to a good performance (in terms
of small Monte Carlo error)  is typically difficult, and may require tedious
pilot
runs. As any tuning parameter, it would be nice to design a strategy
that determines automatically a reasonable value of $N_x$. Second,
\citet{PMCMC} show that, in order to obtain reasonable acceptance
rates for a particle Metropolis-Hastings step, one should take $N_x=\O(t)$, where
$t$ is the number of data-points currently considered. In the \SMCSQ context,
this means that it may make sense to use a small value for $N_x$ for
the early iterations, and then to increase it regularly.  Finally,
when the variance of the PF estimates depends on $\theta$, it might be
interesting to allow $N_x$ to change with $\theta$ as well.

The \SMCSQ
framework provides more scope for such adaptation compared to 
PMCMC. In this section we describe two possibilities, which
relate to the two main particle MCMC
methods, particle marginal Metropolis-Hastings and particle Gibbs. The former 
generates the auxiliary variables independently of the current
particle system whereas the latter does it conditionally on the
current system. For this reason the latter yields a new system without
changing the weights, which is a nice feature, but it requires storing
particle histories, which is memory inefficient; see Section \ref{sub:complex}
for a more thorough discussion of the memory cost of \SMCSQ. 

The schemes for increasing $N_x$ can be integrated into the main
\SMCSQ algorithm along with rules for automatic calibration. We propose the
following simple strategy. We start with a small value 
for $N_x$, we monitor the acceptance rate of the PMCMC
step and
when this rate falls below a given threshold, we trigger
the ``changing $N_x$'' step; for example 
we multiply $N_x$ by 2. 

\subsubsection{Exchange importance sampling step}\label{sub:exchange}

Our first suggestion involves a particle exchange. At iteration $t$, 
the algorithm has generated so far the random variables $\theta^{1:N_\theta}$,
$x_{1:t}^{1:N_x,1:N_\theta}$ and $a_{1:t-1}^{1:N_x,1:N_\theta}$ and  the target distribution is
$\pi_{t}(\theta,x_{1:t}^{1:N_x},a_{1:t-1}^{1:N_x})$.
At this stage, one may extend the sampling space, by generating for each particle
$\theta^m$,  new PFs  of size $\tilde{N}_x$, by simply sampling independently,
for each $m$, the random variables
$\tilde{x}_{1:t}^{1:\tilde{N}_x,m},\tilde{a}_{1:t-1}^{1:\tilde{N}_x,m}$
from $\psi_{t,\theta^m}$. Thus, the extended target distribution
is: 
\begin{equation}
  \label{eq:extended}
  \pi_{t}(\theta,x_{1:t}^{1:N_x},a_{1:t-1}^{1:N_x})\psi_{t,\theta}(\tilde{x}_{1:t}^{
1:\tilde{N}_x},\tilde{a}_{1:t-1}^{1:\tilde{N}_x}).
\end{equation}
In order to swap the $x-$particles and the $\tilde{x}-$particles,
we use the generalised importance sampling strategy of \cite{DelDouJas:SMC},
which is based on an artificial backward kernel. Using \eqref{eq:def_pit},  we compute the incremental 
weights 
\begin{eqnarray}
\label{eq:importanceratio}
\lefteqn{
\frac{\pi_t(\theta,\tilde{x}_{1:t}^{1:\tilde{N}_x},\tilde{a}_{1:t-1}^{1:\tilde{N}_x})
L_t\left( (\theta,\tilde{x}_{1:t}^{1:\tilde{N}_{x}},\tilde{a}_{1:t-1}^{1:\tilde{N}_{x}}),
(x_{1:t}^{1:N_{x}},a_{1:t-1}^{1:N_{x}})\right)}
{\pi_t(\theta,x_{1:t}^{1:N_x},a_{1:t-1}^{1:N_x})\psi_{t,\theta}(\tilde{x}_{1:t}^{1:\tilde{N}_x},\tilde{a}_{1:t-1}^{1:\tilde{N}_x})}} & & \\
 & = & \frac{\hat{Z}_{t}(\theta,\tilde{x}_{1:t}^{1:\tilde{N}_{x}},\tilde{a}_{1:t-1}^{1:\tilde{N}_{x}})}
{\hat{Z}_{t}(\theta,x_{1:t}^{1:N_x},a_{1:t-1}^{1:N_x})}
\times\frac{L_t\left( (\theta,\tilde{x}_{1:t}^{1:\tilde{N}_{x}},\tilde{a}_{1:t-1}^{1:\tilde{N}_{x}}),
(x_{1:t}^{1:N_{x}},a_{1:t-1}^{1:N_{x}})\right)}
{\psi_{t,\theta}(x_{1:t}^{1:N_x},a_{1:t-1}^{1:N_x})}
\nonumber
\end{eqnarray}
where  $L_t$ is a backward kernel density.  
One then may drop  the ``old'' particles $(x_{1:t}^{1:N_x},a_{1:t-1}^{1:N_x})$ in order to 
obtain a new particle system, based on particles $(\theta,\tilde{x}_{1:t}^{1:\tilde{N}_x},\tilde{a}_{1:t-1}^{1:\tilde{N}_x})$
 targetting $\pi_t$, but with $\tilde{N}_x$, $x-$particles.

This importance sampling operation is valid under mild assumptions for
the backward kernel $L_t$; namely that the support of the denominator of \eqref{eq:importanceratio}
is included in the support of its numerator. One easily deduces from  Proposition 1 of \cite{DelDouJas:SMC} and  \eqref{eq:def_pit}
that  the optimal kernel (in terms of minimising the variance of the weights) is 
\begin{equation*}
L_t^{\mathrm{opt}}\left( (\theta,\tilde{x}_{1:t}^{1:\tilde{N}_{x}},\tilde{a}_{1:t-1}^{1:\tilde{N}_{x}}),
(x_{1:t}^{1:N_{x}},a_{1:t-1}^{1:N_{x}})\right)
=\frac{\psi_{t,\theta}(x_{1:t}^{1:N_{x}},a_{1:t-1}^{1:N_{x}})\hat{Z}_{t}
(\theta,x_{1:t}^{1:N_{x}},a_{1:t-1}^{1:N_{x}})}{p(y_{1:t}|\theta)}.
\end{equation*}
This function is intractable, because of the denominator
$p(y_{1:t}|\theta)$, but it 
suggests the following simple approximation: $L_t$ should be set to $\psi_{t,\theta}$, so as to cancel the second ratio,
 which leads to the very simple incremental weight function: 

\begin{equation*} 
u_t^{exch} \left(\theta,x_{1:t}^{1:N_x},a_{1:t-1}^{1:N_x},
\tilde{x}_{1:t}^{1:\tilde{N}_x},\tilde{a}_{1:t-1}^{1:\tilde{N}_x} \right) 
= \frac{\hat{Z}_t(\theta,\tilde{x}_{1:t}^{1:\tilde{N}_x},\tilde{a}_{1:t-1}^{1:\tilde{N}_x})}
{\hat{Z}_t(\theta,x_{1:t}^{1:N_x},a_{1:t-1}^{1:N_x})}.
\end{equation*}

By default, one may implement this exchange step for all the particles $\theta^{1:N_\theta}$, 
and multiply consequently each particle weight $\omega^m$ with the ratio above. 
However, it is possible to apply this step to only a subset
of particles, either selected randomly or according to some deterministic criterion
based on $\theta$. (In that case, only the weights of the selected particles
should be updated.)  Similarly, one could update certain particles according 
to a Hastings-Metropolis step, where the exchange operation is proposed,
and accepted with probabilty the minimum of 1 and the ratio above. 

In both cases, one effectively targets a mixture of $\pi_t$ distributions
corresponding to different values of $N_x$. This does not pose any formal 
difficutly, because these distributions admit the same marginal distributions
with respect to the components of interest ($\theta$, and $x_{1:t}$ if the target
distribution is extended as described in Section \ref{sub:invariant}),
and because the successive importance sampling steps 
(such as either the exchange step above, or Step (b) in the \SMCSQ Algorithm) 
correspond to ratios of densities that are known up to a constant that does not depend on $N_x$. 

Of course, in practice, propagating PF of varying size $N_x$ is a bit more cumbersome
to implement, but it may show useful in particular applications, where for instance
the computational cost of sampling a new state $x_{t+1}$, conditional on $x_t$,
varies strongly according to $\theta$. 

\subsubsection{Conditional SMC step} \label{sub:condsmc}

Whereas the exchange steps associates with the target $\pi_t$, and the particle
Metropolis-Hastings algorithm, our second suggestion  relates to the target 
$\pi_t^\star$, and to the particle Gibbs algorithm. 
First, one extends the target distribution, from $\pi_t$ to $\pi_t^\star$, 
by sampling a particle index $n^\star$, as explained in Section \ref{sub:invariant}. 
Then one may apply a conditional SMC step \citep{PMCMC}, to generate
a new particle filter of size $\tilde{N}_x$, 
$\tilde{x}_{1:t}^{1:\tilde{N}_x},\tilde{a}_{1:t-1}^{1:\tilde{N}_x}$,
but conditional on one trajectory being equal to $\x_{1:t}^n$. This amounts
to sampling the conditional distribution defined by the two factors
in curly brackets in \eqref{eq:Andrieuetal_pit}, which can also be conveniently rewritten as
\begin{equation*}
\frac{ N_x^t\pi_t^\star(n^\star,\theta,\tilde{x}_{1:t}^{1:\tilde{N}_x},\tilde{a}_{1:t-1}^{1:\tilde{N}_x})}
{ p(\theta,\x_{1:t}^n|y_{1:t}) }.
\end{equation*}

We refrain from calling this operation a Gibbs step, because it changes
the target distribution (and in particular its dimension), from $\pi_t(\theta,x_{1:t}^{1:N_x},a_{1:t}^{1:N_x})$ 
to $\pi_t(\theta,x_{1:t}^{1:\tilde{N}_x},a_{1:t}^{1:\tilde{N}_x})$. 
A better formalisation is again in terms of an importance sampling step 
involving a backward kernel \citep{DelDouJas:SMC}, from the proposal distribution, 
the current target distribution $\pi_t$ 
times the conditional distribution of the newly generated variables: 
\begin{equation*}
\pi_t^\star(n^\star,\theta,x_{1:t}^{1:N_x},a_{1:t-1}^{1:N_x})
\frac{ N_x^t\pi_t^\star(n^\star,\theta,\tilde{x}_{1:t}^{1:\tilde{N}_x},\tilde{a}_{1:t-1}^{1:\tilde{N}_x})}
{ p(\theta,\x_{1:t}^n|y_{1:t}) }
\end{equation*}
towards target distribution
\begin{equation*}
\pi_t^\star(n^\star,\theta,\tilde{x}_{1:t}^{1:\tilde{N}_x},\tilde{a}_{1:t-1}^{1:\tilde{N}_x})
L_t\left( 
(n^\star,\theta,\tilde{x}_{1:t}^{1:\tilde{N}_x},\tilde{a}_{1:t-1}^{1:\tilde{N}_x}),
\cdot \right)
\end{equation*}
where $L_t$ is again an arbitrary backward kernel, whose argument, denoted by a dot, is all the variables
in $(x_{1:t}^{1:N_x},a_{1:t-1}^{1:N_x})$, except the variables corresponding to trajectory $\x_{1:t}^n$. 
It is easy to see that the optimal backward kernel (applying again Proposition 1 of \citealt{DelDouJas:SMC})
is such that the importance sampling ratio equals one. 
The main drawback of this approach is that it requires to store all the state variables $x_{1:t}^{1:N_x,1:N_\theta}$;
see our dicussion of memory cost in Section \ref{sub:complex}.   
%
%
%
%
 

%
%
%
%

\subsection{Complexity}
\label{sub:complex}

\subsubsection{Memory cost}

In full generality the  \SMCSQ algorithm is memory-intensive:
up to iteration $t$, $\O(tN_\theta N_x)$ variables have been generated
and potentially have to be carried forward to the next
iteration. We explain now how this cost can be reduced 
to $\O(N_\theta N_x)$  with little loss of generality. 

Only the variables 
$x_{t-1}^{1:N_x,1:N_\theta}$  are necessary to carry out Step (a) of
the algorithm, while all other state variables $x_{1:t-2}^{1:N_x,1:N_\theta}$ can be
discarded. Additionally, when Step  (c) is carried out as described in
Section \ref{sub:PMCMCstep}, $\hat{Z}_t$ is the only additional necessary
statistic of the
particle histories. Thus, the typical implementation of \SMCSQ for
sequential parameter estimation, filtering and prediction has an
$\O(N_\theta N_x)$ memory cost. The memory cost of the exchange step is also $O(N_x N_\theta)$;
more precisely, it is $O(\tilde{N}_x N_\theta)$, where $\tilde{N}_x$ is the new 
size of the PF's. A nice property of 
this exchange step is that it temporarily regenerates complete trajectories $x_{1:t}^{1:\tilde{N}_x,m}$,
sequentially for $m=1,\ldots,M$. Thus, besides augmenting $N_x$ dynamically, 
the exchange step can also be used to to carry out operations involving 
complete trajectories at certain pre-defined times, 
while maintaining a $\O(N_\theta \tilde{N}_x)$ overall cost. 
Such operations include  inference
with respect to $\pi_t(\theta,x_{1:t})$, updating $\theta$ with respect to 
the full conditional $p(\theta|x_{1:t},y_{1:t})$, as explained in Section \ref{sub:invariant},
or even the conditional SMC update descrided in Section
\ref{sub:condsmc}. 

\subsubsection{Stability and computational cost}

 Step (c), which requires  re-estimating  the
likelihood, is the most computationally expensive component of \SMCSQ. When this
operation is performed at time $t$, it incurs an $\O(tN_\theta N_x)$
computational cost. Therefore, to study the computational cost of
\SMCSQ we need to investigate  the rate at which ESS drops below a given
threshold. This question directly relates to the stability  of the filter, and we
will work as in 
Section 3.1 of \cite{Chopin:CLT} to answer it. Our approach is  based on
certain simplifying assumptions, regularity conditions and a recent
result of  \cite{CerDelGuy2011nonasymptotic} which all lead
to Proposition  \ref{lem:comp}; the assumptions are discussed in some detail in
Appendix B. 

In general, $\mathrm{ESS}/N_\theta < \gamma$, for ESS given in
\eqref{eq:ess}, is a standard degeneracy criterion of sequential
importance sampling due to the fact that the limit of $\mathrm{ESS}/N_\theta$ as
$N_\theta \to \infty$  is  equal to the inverse
of the second moment of the importance sampling weights (normalized to
have mean 1). This limiting quantity,  which we will generically denote by
$\mathcal{E}$,  is also often called 
effective sample size since  it can be 
interpreted as an equivalent number of 
independent samples
from the target distribution \citep[see  Section
2.5.3 of ][for details]{liu:book}.  The first simplification in our
analysis is to study the properties of $\mathcal{E}$, rather than its
finite sample estimator $\textrm{ESS}/N_\theta$, and consider an algorithm which resamples
whenever  $\mathcal{E}<\gamma$.  

Consider now the specific context of  \SMCSQ. Let $t$ be a resampling
time at which equally weighted, independent
particles have been obtained, and  $t+p,\,p>0$, a future time such
that no resampling has happened since $t$. The marginal distribution of the resampled particles at time $t$ 
 is only approximately  $\pi_t$ due to the 
burn-in period of the Markov chains which are used to generate them.   The second
simplifying assumption in our analysis is that this marginal distribution is
precisely $\pi_t$.  Under this assumption, 
the particles at
time $t+p$ are generated according to the distribution
$\bar{\pi}_{t,t+p}$, 
\begin{equation*}
 \bar{\pi}_{t,t+p}(\theta,x_{1:t+p}^{1:N_{x}},a_{1:t+p-1}^{1:N_{x}}) 
  =   
\pi_t(\theta, x_{1:t}^{1:N_{x}},a_{1:t-1}^{1:N_{x}}) \,
\psi_{t+p,\theta}(x_{1:t+p}^{1:N_{x}},a_{1:t+p-1}^{1:N_{x}} \mid
x_{1:t}^{1:N_{x}},a_{1:t-1}^{1:N_{x}}) 
\end{equation*}
and the expected value of the  weights $\omega_{t+p}$ obtained from \eqref{eq:ishat} is
$p(y_{1:t})/p(y_{1:t+p})$. Therefore, the normalized weights are given
by 
$$
\frac{p(y_{1:t})}{p(y_{1:t+p})} \prod_{i=1}^{p}
\hat{p}(y_{t+i}|y_{1:t+i-1},\theta) 
  \stackrel{\Delta}{=} 
 \frac{\hat{Z}_{t+p|t}(\theta,x_{1:t+p}^{1:N_{x}},a_{1:t+p-1}^{1:N_{x}})}{p(y_{t+1:t+p}|y_{1:t})}\,,
$$
and  the inverse of the  second moment of the
normalized weights in \SMCSQ and  IBIS is given by
$$
\mathcal{E}_{t,t+p}^{N_{x}} =  \left \{
  \mathbb{E}_{\bar{\pi}_{t,t+p}}\left[\frac{\hat{Z}_{t+p|t}(\theta,x_{1:t+p}^{1:N_{x}},a_{1:t+p-1}^{1:N_{x}})^2}{p(y_{t+1:t+p}|y_{1:t})^2}
    \right]
\right \}^{-1} , \quad \mathcal{E}_{t,t+p}^{\infty} =  \left \{
  \mathbb{E}_{p(\theta|y_{1:t})}\left[{p(\theta|y_{1:t+p})^2 \over
      p(\theta|y_{1:t})^2} \right]
\right \}^{-1}\,.
$$
The previous development leads to the following Proposition which is proved in Appendix B.
\begin{lem}
\label{lem:comp}
\begin{enumerate}
\item Under Assumptions (H1a) and (H1b)  in Appendix B,  there exists
  a constant $\eta>0$ such that for any $p$, if $N_x
  > \eta p$,  
  \begin{equation}
    \label{eq:ess-order}
   \mathcal{E}_{t,t+p}^{N_{x}}  \geq {1 \over 2}  \mathcal{E}_{t,t+p}^{\infty}.
  \end{equation}
\item Under Assumptions (H2a)-(H2d) in Appendix B, for any $\gamma>0$
  there exist  $\tau,\eta>0$ and $t_0<\infty$, such that  for $t\geq t_{0}$,
\[
\mathcal{E}_{t,t+p}^{N_{x}}\geq \gamma,\mbox{ for }p=\left\lceil \tau t\right\rceil ,\, N_{x}=\left\lceil \eta t\right\rceil .
\]
\end{enumerate}
\end{lem}

The implication of this Proposition is the following: under the
assumptions in  Appendix B and the assumption that the resampling step
produces samples from the target distribution, the resample steps should be triggered at times $\left\lceil \tau^{k}\right\rceil $,
$k\geq1$, to ensure that the weight degeneracy between two successive
resampling step stays bounded in the run of the algorithm; at these
times
$N_{x}$ should be adjusted to $N_{x}=\left\lceil \eta \tau^k \right\rceil $;
thus, the cost of each successive importance sampling step is $\O(N_{\theta}\tau^{k})$,
until the next resampling step; a simple calculation shows that the
cumulative computational
cost of the algorithm up to some iteration $t$ is then
$\O(N_{\theta}t^{2})$. This is to be contrasted with a computational
cost $\O(N_{\theta}t)$ for IBIS  under a similar  set of
assumptions. The assumptions which lead to this result are restrictive
but they are typical of the state of the art for obtaining
results about the stability of this type of sequential algorithms; see
Appendix B for further discussion.

%
%
%
%


\section{Numerical illustrations}\label{sec:numerics}

An initial study which illustrates \SMCSQ in a range of examples 
of moderate difficulty is available from the second author's web-page, 
see \href{http://sites.google.com/site/pierrejacob/}{http://sites.google.com/site/pierrejacob/}, 
as supplementary material.  
In that study, \SMCSQ was shown to 
typically outperform competing algorithms, whether in sequential
scenarios (where datapoints are obtained sequentially)
or in batch scenarios (where the only distribution of interest
is $p(\theta,x_{1:T}|y_{1:T})$ for some fixed time horizon $T$). 
For instance, in the former case,  \SMCSQ was shown to provide  smaller Monte Carlo errors
than the SOPF at a given CPU cost. In the latter case, 
\SMCSQ was shown to compare favourably to an adaptive version of the marginal
PMCMC algorithm proposed by \citet{PetersHosackHayes}. 

In this paper, our objective instead is to take a hammer to \SMCSQ, 
that is, to evaluate its performance on models that are regarded
as particularly challenging, even for batch estimation purposes. 
In addition, we treat \SMCSQ as much as possible as a black box: 
the number $N_x$ of $x$-particles is augmented dynamically (using the exchange step, see
Section \ref{sub:exchange}), as explained
in Section \ref{sec:increase}; the move steps are calibrated using the current particles, as 
described at the end of Section \ref{sub:An-idealized-algorithm}, and so on.  
The only model-dependent inputs are (a) a procedure
for sampling from the Markov transition of the model, $f_\theta(x_{t+1}|x_t)$;
(b) a procedure for pointwise evaluation the likelihood $g_\theta(y_t|x_t)$; 
and (c) a prior distribution on the parameters. 
This means that the proposal $q_{t,\theta}$ is set to the default
choice $f_\theta(x_{t+1}|x_t)$. This also means that we are able to treat 
models such that the density  $f_\theta(x_{t+1}|x_t)$ cannot be 
computed, even if it may be sampled from; this is the case
in the first application we consider. 

A generic \SMCSQ software package written in Python and C by the second author is available at 
\href{http://code.google.com/p/py-smc2/}{http://code.google.com/p/py-smc2/}. 

\subsection{Sequential prediction of asset price volatility}

\SMCSQ  is particularly well suited to tackle several of the
challenges that arise in the probabilistic modelling of financial time
series: prediction is of central importance; risk management requires
accounting for  parameter and model uncertainty; 
non-linear models are necessary to capture the features in the data; the length of typical time series is large when
modelling 
medium/low frequency data and vast when considering 
high frequency observations.

We illustrate some of these possibilities in the context of prediction
of daily volatility of asset prices. There is a vast literature on
stochastic volatility (SV) models; we simply refer to the excellent
exposition in  \cite{bns:real} for references, perspectives and
second-order properties. The generic framework for daily
volatility is as 
follows. Let $s_t$ be the value of a given
financial asset (e.g a stock price or an exchange rate) on the $t$-th
day, and $y_t=10^{5/2} \, \log(s_t/s_{t-1})$ be the so-called log-returns (the
scaling is done for numerical convenience). 
The SV model specifies a state-space model
with observation  equation: 
\begin{equation}
  \label{eq:sv-obs}
  y_t = \mu + \beta v_t + v_t^{1/2} \epsilon_t \,, t\geq 1
\end{equation}
where the $\epsilon_t$ is a sequence of independent errors which are
assumed to be standard Gaussian. The process $v_t$ is known as the actual
volatility and  it is treated as a stationary stochastic
process. This implies that log-returns are stationary with mixed Gaussian
marginal distribution. The coefficient $\beta$ has both a financial
interpretation, as a risk premium for excess volatility, and a
statistical one, since  for $\beta \neq 0$ the marginal density of
log-returns is skewed. 

We consider the class of L\'evy driven SV
models which were introduced in \cite{bns:ou} and have been
intensively studied in the last decade from both the mathematical
finance and the statistical community. 
This family of models is specified via a  continuous-time model for the joint
evolution of log-price and spot (instantaneous)
volatility, which are driven by
Brownian motion and L\'evy process respectively. The actual volatility
is the integral of the spot volatility over daily intervals, and the
continuous-time model  translates into
a state-space model for $y_t$ and $v_t$ as we show below.   Details can be
found in Sections 2 (for the continuous-time specification) and 5
(for the state-space
representation) of the original article.
 Likelihood-based inference for this class of models is recognized
 as a very challenging problem, and it 
 has been
undertaken among others in \cite{robe:papa:dell:2004,grif:steel:ou}
and most recently in \cite{PMCMC} using PMCMC. On the other hand,
\cite{bns:real} develop quasi-likelihood methods using the Kalman filter based on an
approximate state-space formulation suggested by the second-order properties
of the $(y_t,v_t)$ process.  

Here we focus on models where the background
driving L\'evy process  is expressed in terms of a finite rate Poisson process and consider 
multi-factor specifications of such models which include
leverage. This choice allows the exact simulation of the actual
volatility process, and permits direct comparisons to  the numerical
results  in Sections 4 of \cite{robe:papa:dell:2004}, 3.2 of 
\cite{bns:real} and 6 of \cite{grif:steel:ou}. Additionally, this case
is representative  of a system which can be very easily simulated
forwards whereas computation of its transition density is considerably
involved (see \eqref{eq:ssf} below). The specification for the
one-factor model is as follows. We parametrize the latent process as in
\cite{bns:real} in terms of $(\xi,\omega^2,\lambda)$ where $\xi$ and $\omega^2$ are
the stationary mean and variance of the spot volatility process, and
$\lambda$ the exponential rate of decay of its 
autocorrelation function.  The second-order properties of $v_t$ can be
expressed as functions of these parameters,
see Section 2.2 of \cite{bns:real}. The
state dynamics for the actual volatility are as follows: 
\begin{equation}
  \label{eq:ssf}
  \begin{split}
     k  & \sim  \mathrm{Poi}\left( \lambda \xi^2/\omega^2 \right)\,, \quad 
 c_{1:k}  \iid  \mathrm{U}(t,t+1)\,, \quad 
e_{1:k}  \iid  \mathrm{Exp}\left(\xi/\omega^2 \right),  \\ 
  z_{t+1}  & =  e^{-\lambda} z_t + \sum_{j=1}^k  e^{-\lambda(t+1-c_j)}
  e_j\,, \quad
v_{t+1} =  {\frac{1}{\lambda}} \left [ z_t - z_{t+1} + \sum_{j=1}^k e_j
\right ]\,, \quad
x_{t+1} =  (v_{t+1},z_{t+1})' \, .
  \end{split}
\end{equation}
In this representation, $z_t$ is the discretely-sampled spot
volatility process, and the Markovian representation of the state
process involves the pair $(v_t,z_t)$. The random variables $(k,c_{1:k},e_{1:k})$ are generated
independently for each time period, and $1:k$ is understood as the
empty set when  $k=0$. These system dynamics imply a
$\Gamma(\xi^2/\omega^2,\xi/\omega^2)$ as stationary distribution for $z_t$. 
Therefore, we take this to be the initial distribution for $z_0$.

\begin{figure}[htbp]
 \centering
\subfigure[]{\includegraphics[width =
0.3\textwidth]{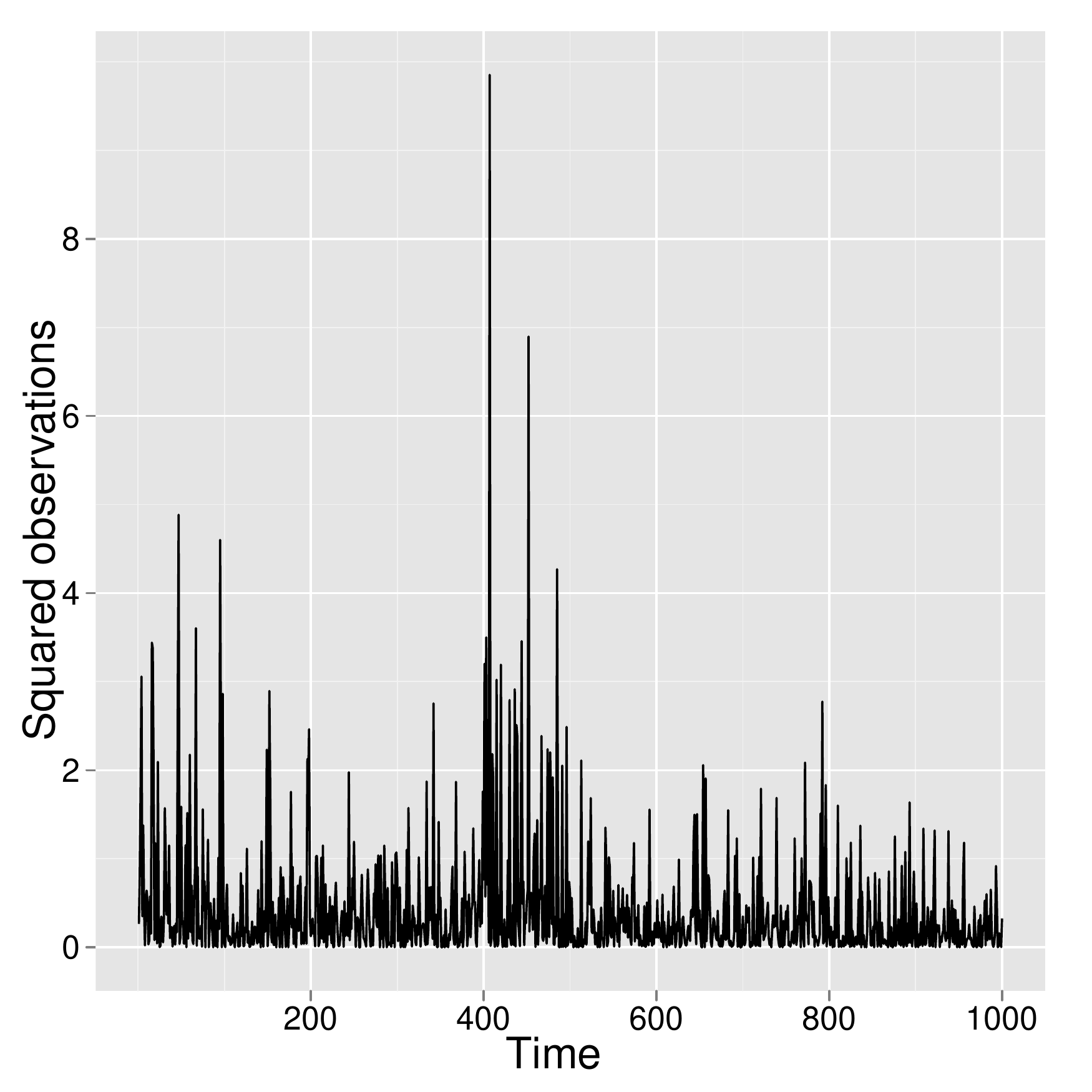}}
 \subfigure[]{\includegraphics[width = 0.3\textwidth]{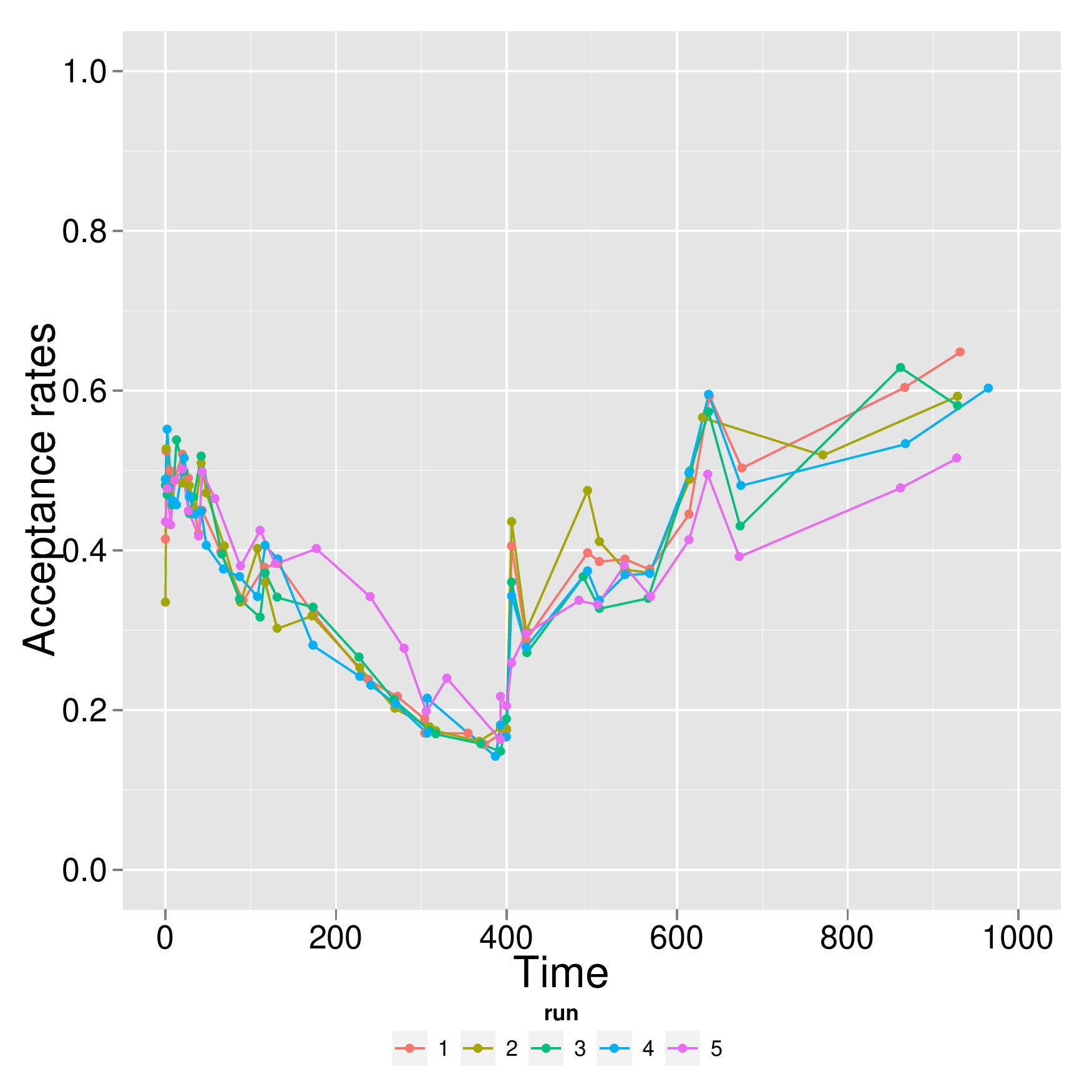}}
 \subfigure[]{\includegraphics[width = 0.3\textwidth]{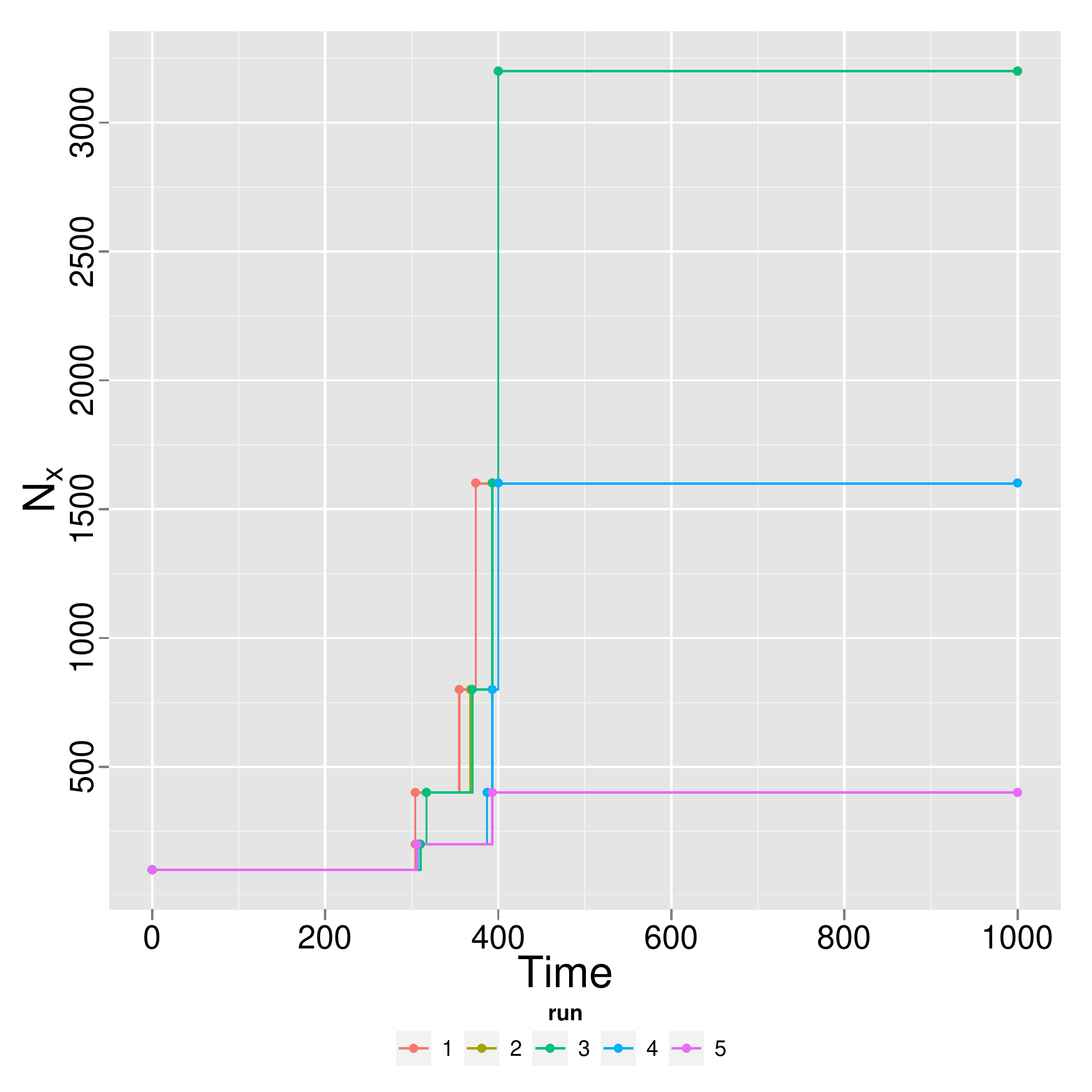}}
 \subfigure[]{\includegraphics[width =
0.90\textwidth]{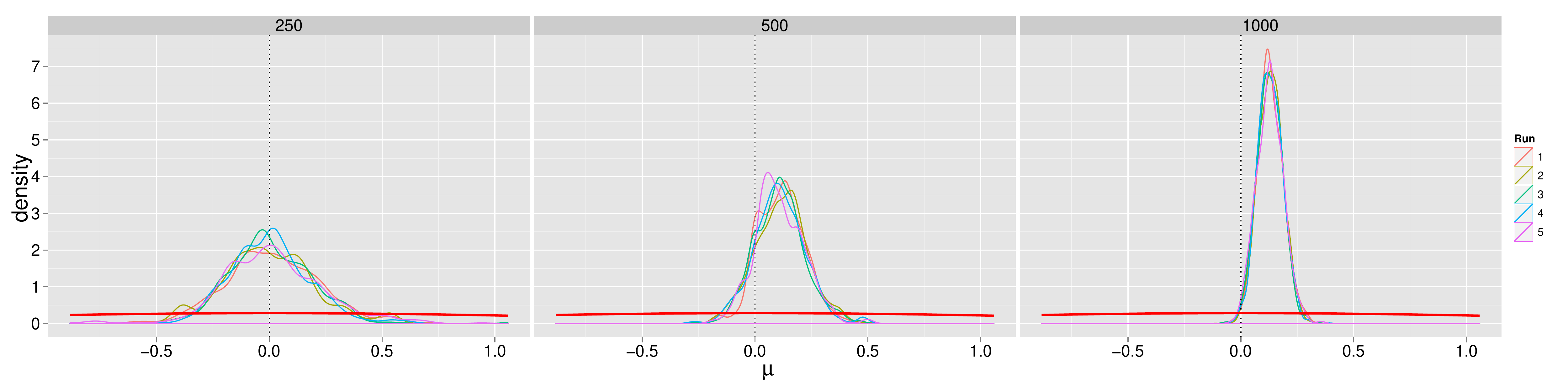}}
 \subfigure[]{\includegraphics[width =
0.90\textwidth]{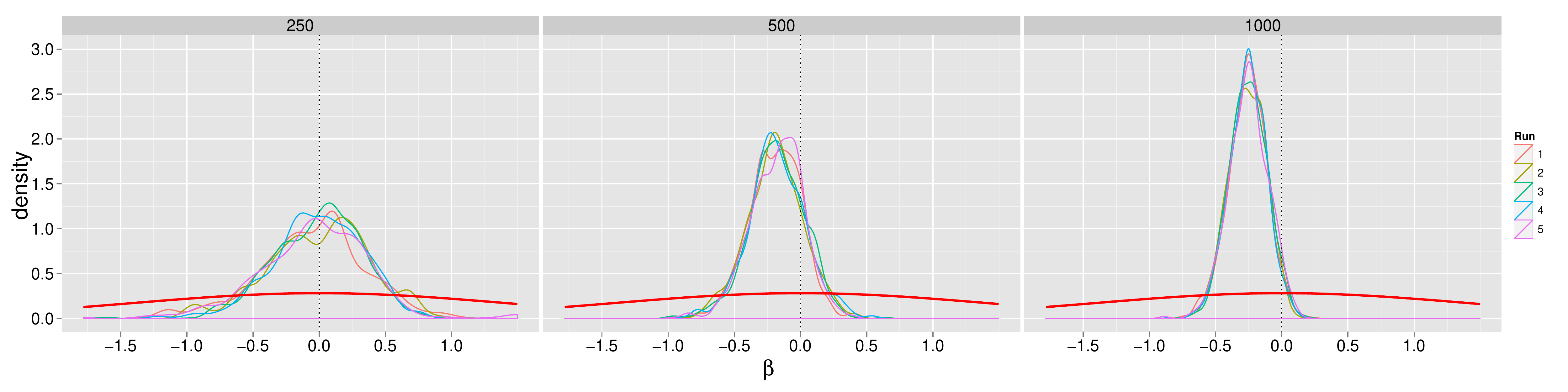}}
 \subfigure[]{\includegraphics[width =
0.90\textwidth]{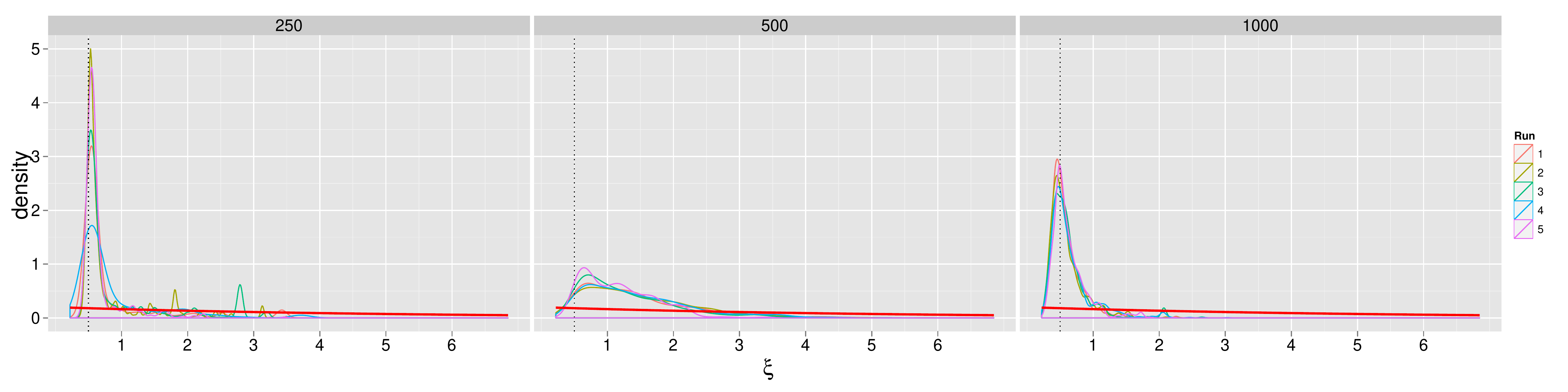}}
 \caption{\label{fig:SVonefactor:Concentration} 
 Single-factor stochastic volatility model, synthetic dataset. (a)
 Squared observations vs time. (b)-(f) Results  obtained from 5
 repeated runs: (b) acceptance rate; (c) $N_x$  vs time; 
(d) to (f) 
overlaid kernel density estimators of the posterior distribution of
$\mu$, $\beta$, $\xi$ 
at different times $t=250,500,1000$, the vertical dashed line indicates the true value and solid red lines
the prior density. }
\end{figure}

We applied the algorithm to a synthetic data set of length $T =
1,000$ (Figure \ref{fig:SVonefactor:Concentration}(a)) simulated with the values
$\mu = 0$, $\beta = 0$, $\xi = 0.5$, $\omega^2 = 0.0625$,
$\lambda = 0.01$  which were used also in the
simulation study of \cite{bns:real}. 
We launched 5 independent runs using $N_\theta = 1,000$, a 
ESS threshold set at $50\%$, and the independent Hastings-Metropolis scheme
described in Section \ref{sub:An-idealized-algorithm}. The number $N_x$ was set initially to 100, and 
increased whenever the acceptance rate went below $20\%$ (Figure \ref{fig:SVonefactor:Concentration}(b)-(c)).
 Figure 
\ref{fig:SVonefactor:Concentration}(d)-(e) shows estimates of the 
 posterior marginal distribution of some parameters. Note  the impact the large jump in the volatility has on 
$N_x$, which is systematically (across runs) increased around time
$400$, and the posterior distribution of the  parameters of the
volatility process, see Figure
\ref{fig:SVonefactor:Concentration}(f).

\begin{figure}[htbp]
  \centering
\subfigure[]{\includegraphics[width = 0.7 \textwidth,
  height=4cm]{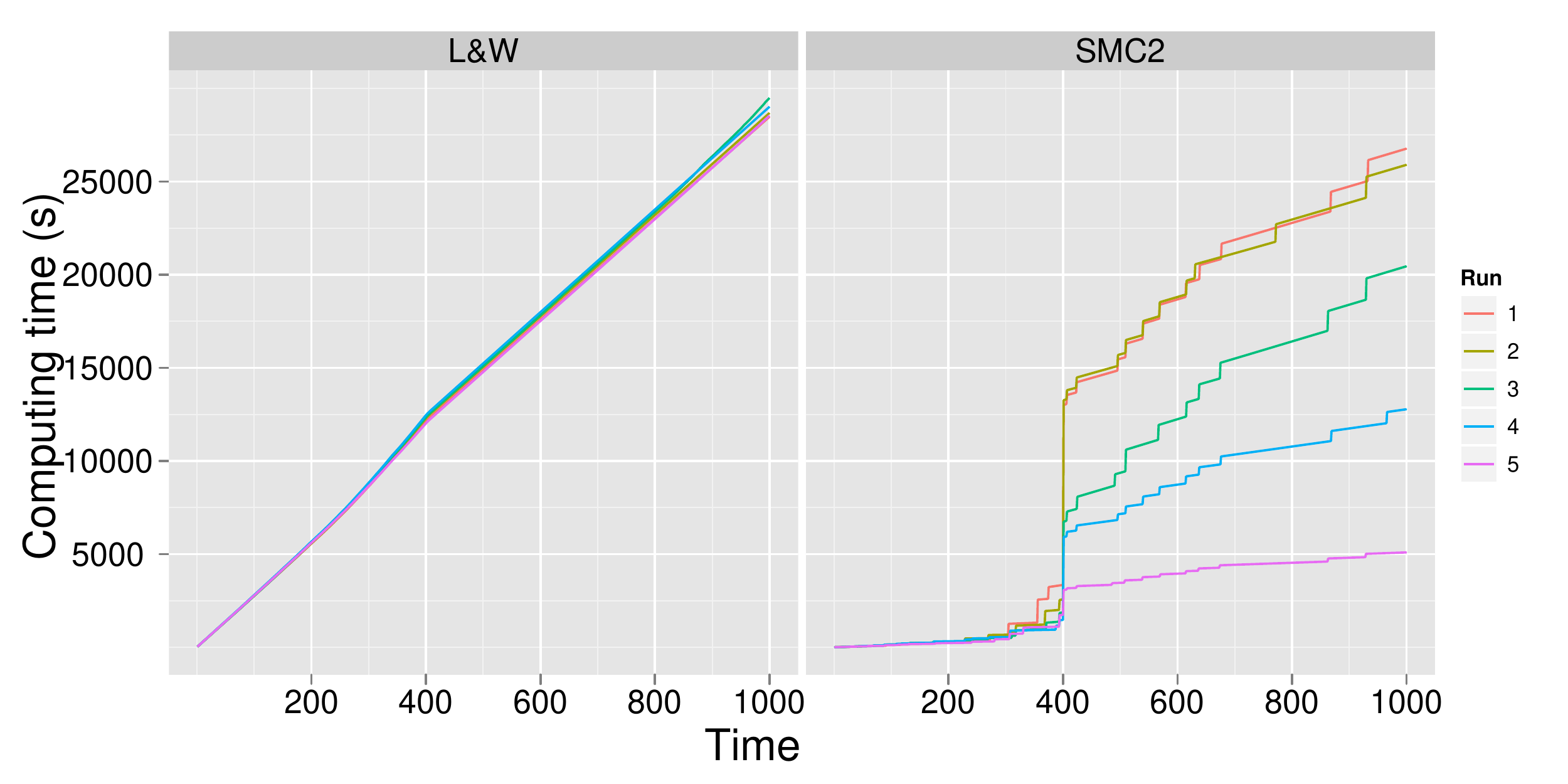}}
\subfigure[]{\includegraphics[width =0.9
  \textwidth]{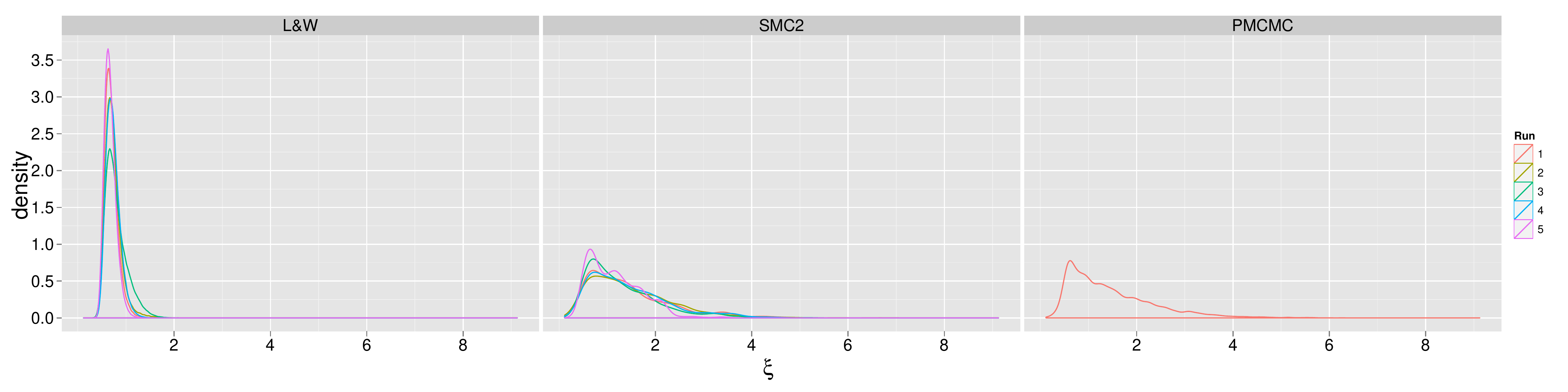}}
\subfigure[]{\includegraphics[width = 0.9 \textwidth, height=4cm]{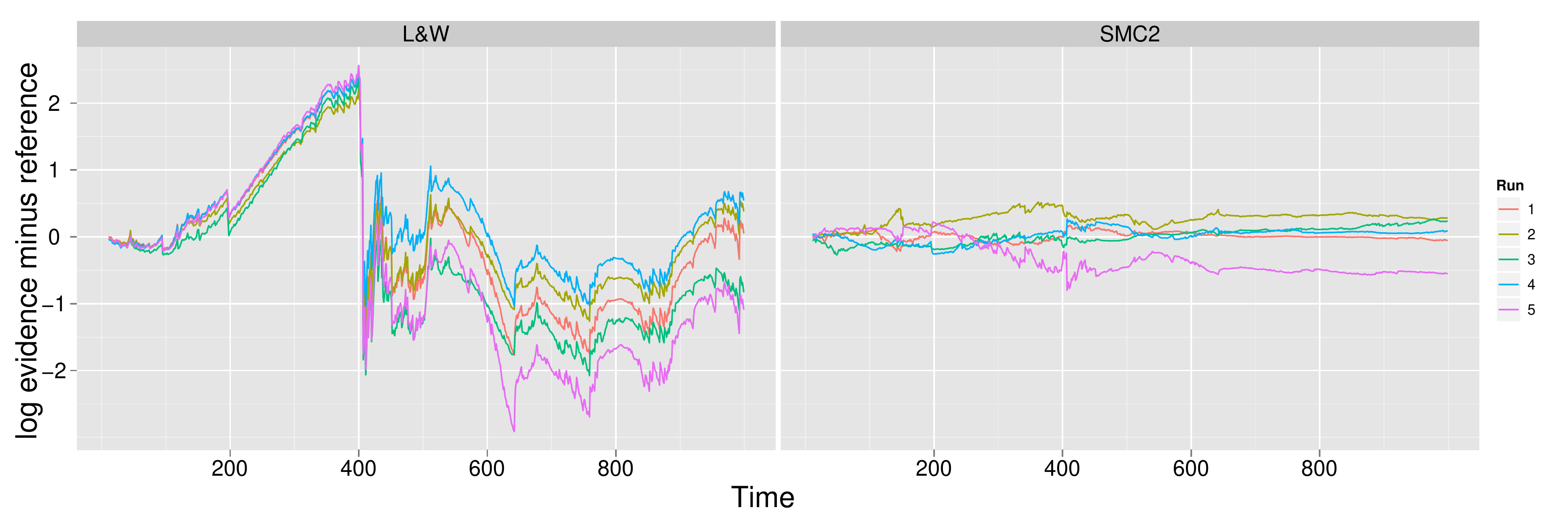}}
  \caption{\label{fig:SVonefactor:CT} Single-factor stochastic
    volatility model, synthetic dataset, comparison between
    methods. (a) Computing time  of $5$ independent runs of L\&W
    (left) and \SMCSQ (right) in seconds, against time. (b)  Estimation of the posterior
marginal distribution of mean volatility, $\xi$. (c) Estimation of the log evidence, the curves represent 
the estimated evidence of each run minus the
mean across $5$ runs of the log evidence estimated using \SMCSQ. }
\end{figure}

It is interesting to compare the numerical performance of \SMCSQ to that of the SOPF and 	
\cite{LiuWest}'s particle filter (referred to as \LW in the
following) for this model and data, and for a comparable CPU budget. 
The SOPF, if run with $N=10^5$ particles, 
collapses to one single particle at about $t=700$ and is thus
completely unusable in this context. \LW is a version of SOPF where
the $\theta$-components of the particles are
diversified using a Gaussian move that leaves the first two empirical
moments of the particle sample unchanged.
This move unfortunately introduces a bias which is hard to quantity. 
We implemented \LW with $N = 2\times 10^5$ $(x,\theta)$-particles and we set the smoothing
parameter $h$ to $10^{-1}$; see the Supplement for results with
various values of $h$. This number of particles was to chosen to make
the computing time of \SMCSQ and \LW comparable,  see
Figure \ref{fig:SVonefactor:CT}(a).  Unsurprisingly, \LW
runs are very consistent in terms of computing times,
whereas those of 
\SMCSQ are more variable, mainly because the number of
$x$-particles does not reach the same value across the runs and  the
number of resample-move steps varies. Each of these runs took between $1.5$ and 
$7$ hours using a simple Python script and only one processing unit of
a 2008 desktop computer (equipped with an Intel Core 2 Duo
E8400). Note that, given that these methods could easily be parallelized, the
computational cost can be greatly reduced; a $100\times$ speed-up is
plausible using appropriate hardware. 

Our results suggest that the bias in \LW is
significant. Figure \ref{fig:SVonefactor:CT}(b) shows the posterior distribution of $\xi$,
the mean of volatility, at time $t = 500$, which is about $100$ time
steps after the large jump in volatility at time $t = 407$. The
results for both algorithms are compared to those from a  long PMCMC
run  \citep[implemented as in ][and detailed in the
Supplement]{PetersHosackHayes} with $N_x = 500$ and $10^5$
iterations. Figure \ref{fig:SVonefactor:CT}(c) reports on  the estimation of the log evidence
$\log p(y_{1:t})$ for each algorithm,  plotting the estimated log
evidence of each run  
minus the mean of the log evidence of the $5$ \SMCSQ runs. 
We see that the log evidence estimated using L\&W is systematically
biased, positively or negatively depending on the time steps, with a
large discontinuity at time $t = 407$, which is due to underestimation
 of the tails of the predictive distribution. 

\begin{figure}[htbp]
 \centering
 \subfigure[]{\label{subfig:SP500:observations2}
\includegraphics [ width =
0.3\textwidth]{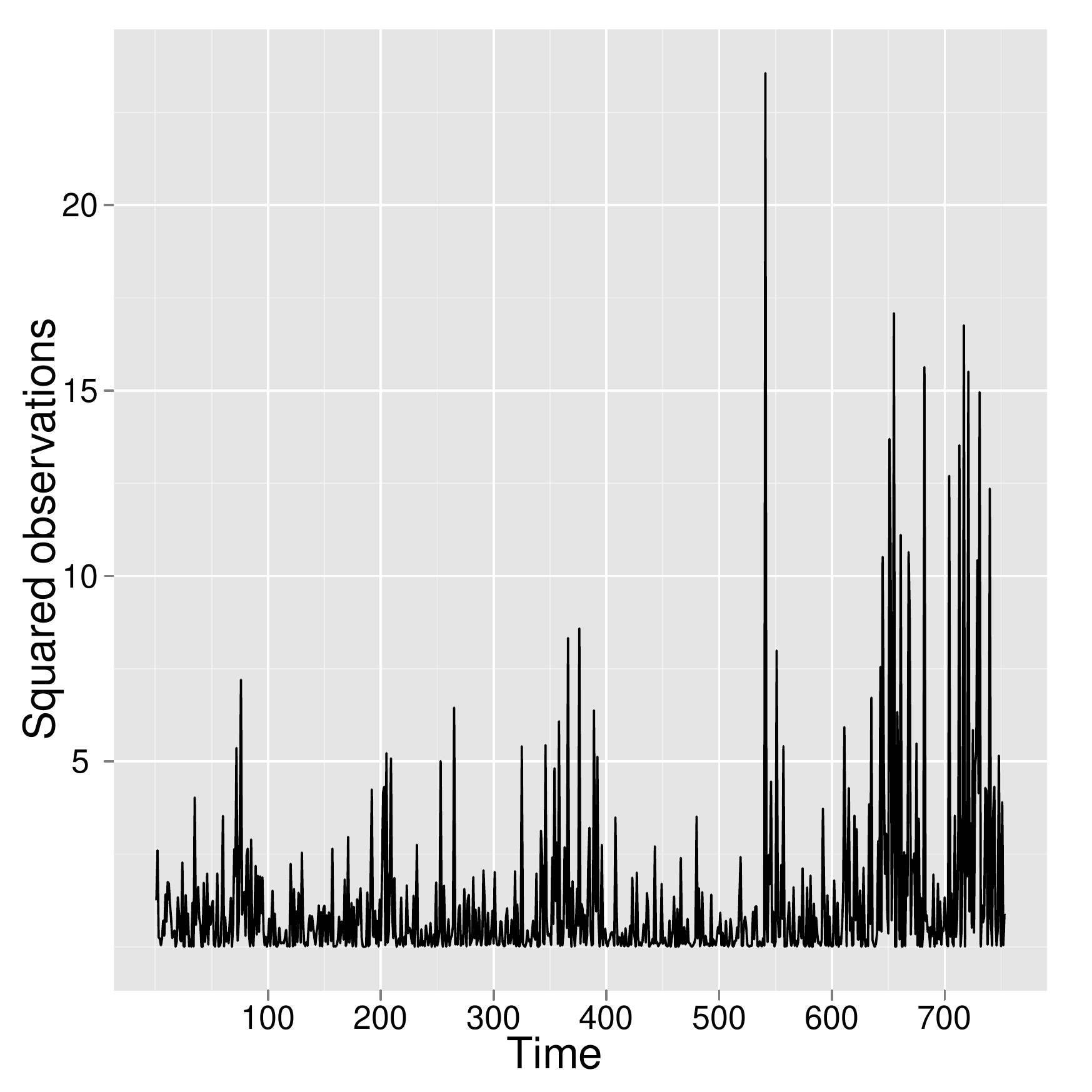}}
 \subfigure[]{\label{subfig:SP500:acceptrates:multifactor}\includegraphics[width
=
0.3\textwidth]{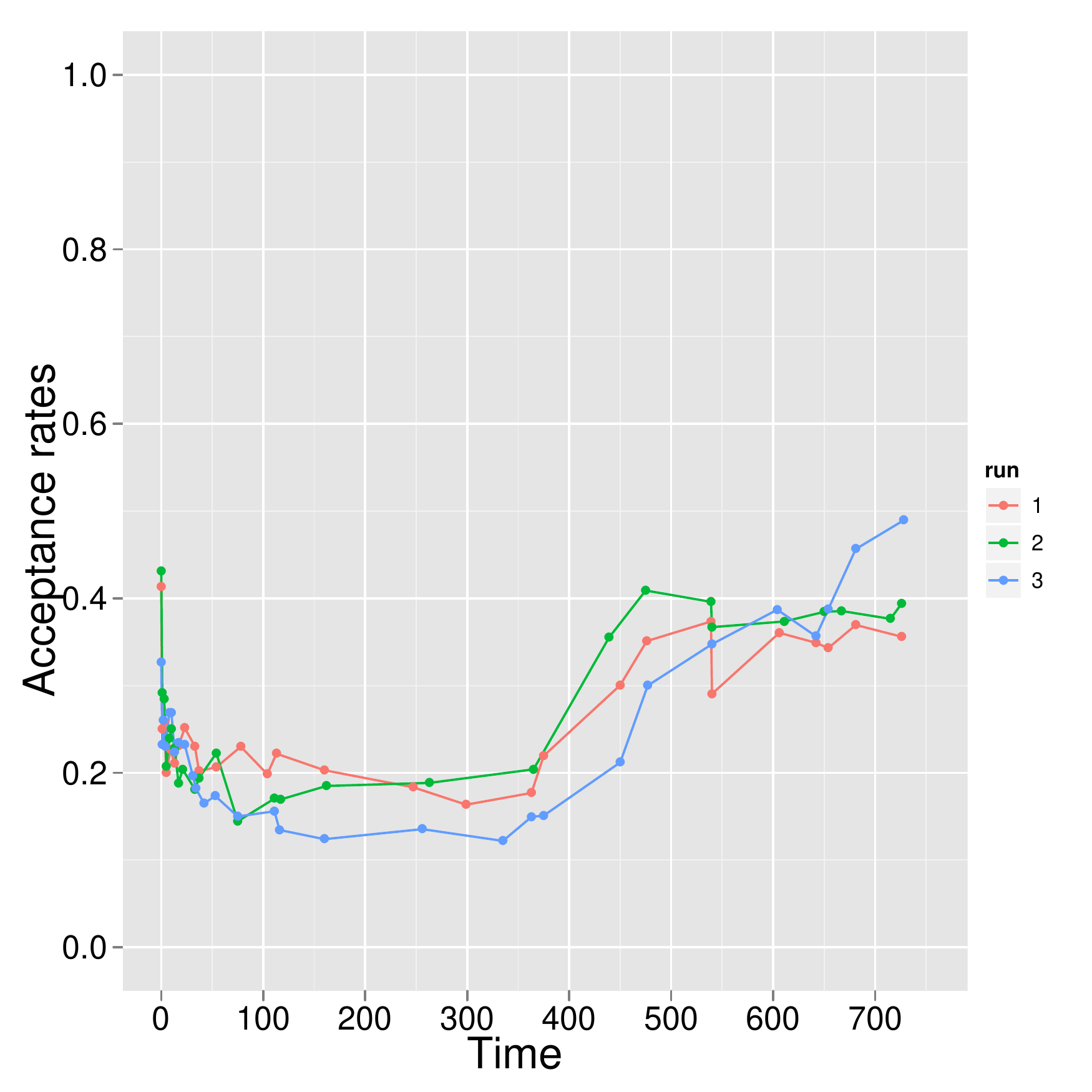}}
 \subfigure[]{\label{subfig:SP500:evidence}\includegraphics[width
=
0.3\textwidth]{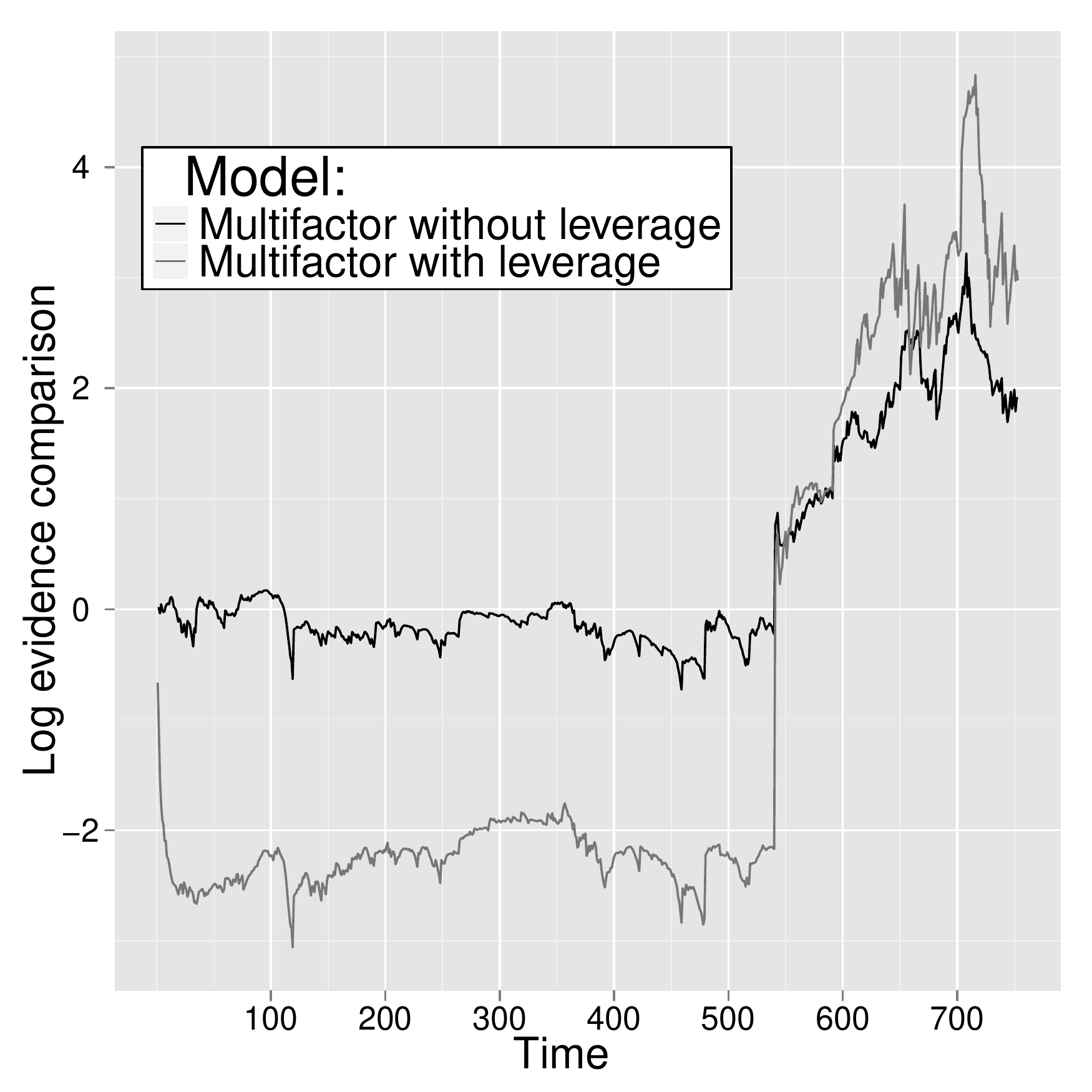}}
 \caption{\label{fig:SP500:observations} (a): the squares of the S\&P 500 data from
   03/01/2005 to 21/12/2007. (b): acceptance
   rates of the resample-move step for the full model over two
   independent runs. (c):   log-evidence comparison between
   models (relative to the one-factor model). }
\end{figure}

We now consider models of different complexity for
the S\&P 500 index. The data set 
is made of $753$ observations from January 3rd 2005 to December 31st
2007 and it is shown on  Figure
\ref{fig:SP500:observations}(a). 
We first consider a two-factor model, according to which the
actual volatility is a sum of two
independent components each of which follows a L\'evy driven
model.  Previous research
indicates that a two-factor model is sufficiently flexible, whereas
more factors do not add significantly when considering daily data, see
for example \cite{bns:real,grif:steel:ou} for L\'evy driven models and
\cite{chernov:multi} for diffusion-driven SV models.  We consider one
component which  describes long-term movements in the
volatility, with memory parameter $\lambda_1$, and another which
captures short-term variation,  with parameter 
$\lambda_2 >> \lambda_1$. The second component allows more freedom in
modelling the tails of the distribution of log-returns. The contribution of the slowly mixing
process to the overall mean and variance of the spot volatility is
controlled by the parameter $w \in (0,1)$. 
Thus, for this model
$x_t=(v_{1,t},z_{1,t},v_{2,t},z_{2,t})$ with
$v_t=v_{1,t}+v_{2,t}$, where each pair $(v_{i,t},z_{i,t})$ evolves
according to  \eqref{eq:ssf} with parameters $(w_i \xi,w_i 
\omega^2,\lambda_i)$ with $w_1=w, w_2=1-w$. The system errors are 
generated by independent sets of variables
$(k_i,c_{i,1:k},e_{i,1:k})$, and $z_{0,i}$ are initialized according
to the corresponding gamma distributions. Finally, we extend the observation equation to capture a significant
feature observed in returns on stocks: low returns provoke
increase in subsequent volatility, see for example \cite{black} for
an early reference. In parameter driven SV models, one generic
strategy to incorporate such feedback is to correlate the noise in the
observation and state processes, see \cite{harvey:shep}  in the
context of the logarithmic  SV model, and Section 3 of \cite{bns:ou}  for L\'evy
driven models.  We take up their suggestion, and re-write the
observation equation as 
\begin{equation}
  \label{eq:sv-obs-lev}
  y_t = \mu + \beta v_t + v_t^{1/2} \epsilon_t + \rho_1
  \sum_{j=1}^{k_1} e_{1,j} + \rho_2  \sum_{j=1}^{k_2} e_{2,j} -
\xi(w \rho_1  \lambda_1 + (1-w) \rho_2  \lambda_2) 
\end{equation}
where $e_{i,j}$ are the system error variables involved in the
generation of $v_t$ and $\rho_i$ are the leverage parameters which we
expect to be negative. Thus, in this specification we deal with a
model with a 5-dimensional state and 9 parameters. 

The mathematical tractability of this family of models and the
specification in terms of stationary and memory parameters allows 
to a certain extent subjective Bayesian modelling. Nevertheless, since
the main emphasis here is to evaluate the performance of \SMCSQ we
choose priors that (as we verify a posteriori) are rather flat in the areas of
high posterior density. Note that   the prior for $\xi$ and $\omega^2$ has to
reflect the scaling of the log-returns by a multiplicative factor. We
take an $\mathrm{Exp}(1)$  prior for $\lambda_1$, 
an $\mathrm{Exp}(0.5)$ for $\lambda_2-\lambda_1$, thus imposing the identifiability
constraint  $\lambda_2>\lambda_1$. We take a $\mathrm{U}(0,1)$ prior for $w$,
an $\mathrm{Exp}(1/5)$ for $\xi$ and $\omega^2$,  and Gaussian priors with large
variances  for the
observation equation parameters.

We launch the three models for the S\&P 500 data: single factor,
multifactor without and with leverage; note that multifactor without leverage 
means the full model, but with $\rho_1=\rho_2=0$ in \eqref{eq:sv-obs-lev}.   We use
 $N_\theta = 2000$, and $N_x$ is set initially to $100$ and then dynamically increases
as already described. 
The acceptance rates stay reasonable as illustrated on Figure
\ref{subfig:SP500:acceptrates:multifactor}. 
Figure \ref{subfig:SP500:evidence} shows the log evidence $\log
p(y_{1:t})$ for the two factor models minus the log evidence for the single
factor model. Negative values at time $t$ means that the observations favour the
single factor model up to time $t$. Notice how the model evidence
changes after the big jump in volatility  around time $t =
550$. Estimated posterior densities for all parameters are provided in the
Supplement.

  \subsection{Assessing extreme athletic records}

\begin{figure}
 \centering
 \subfigure[]{\label{fig:athletics:observations}\includegraphics[
width = 0.45\textwidth, height=4cm]{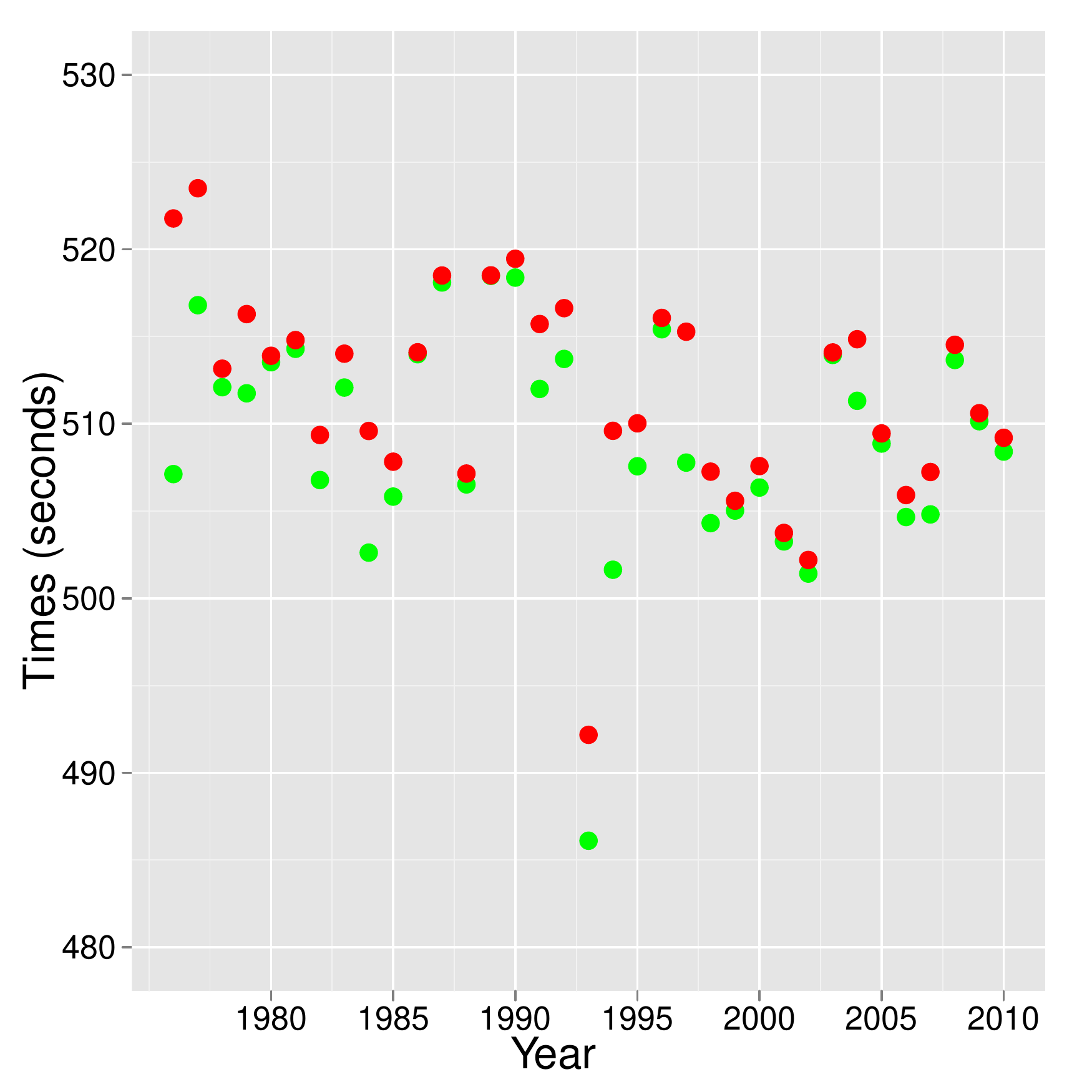}}
 \subfigure[]{\label{fig:athletics:probabeating}\includegraphics[width
=
0.45\textwidth, height=4cm]{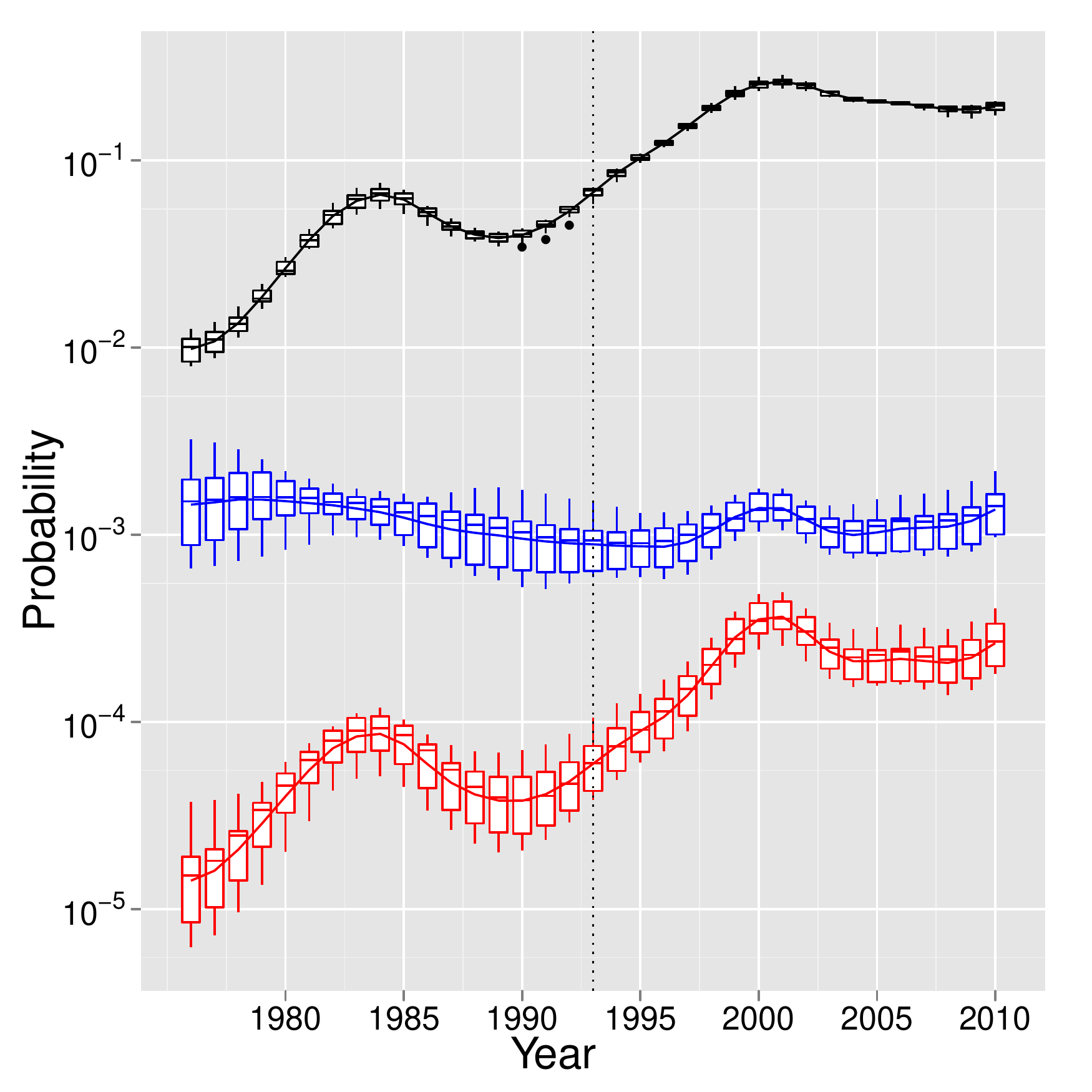}}
 \subfigure[]{\includegraphics[width = 0.3\textwidth]{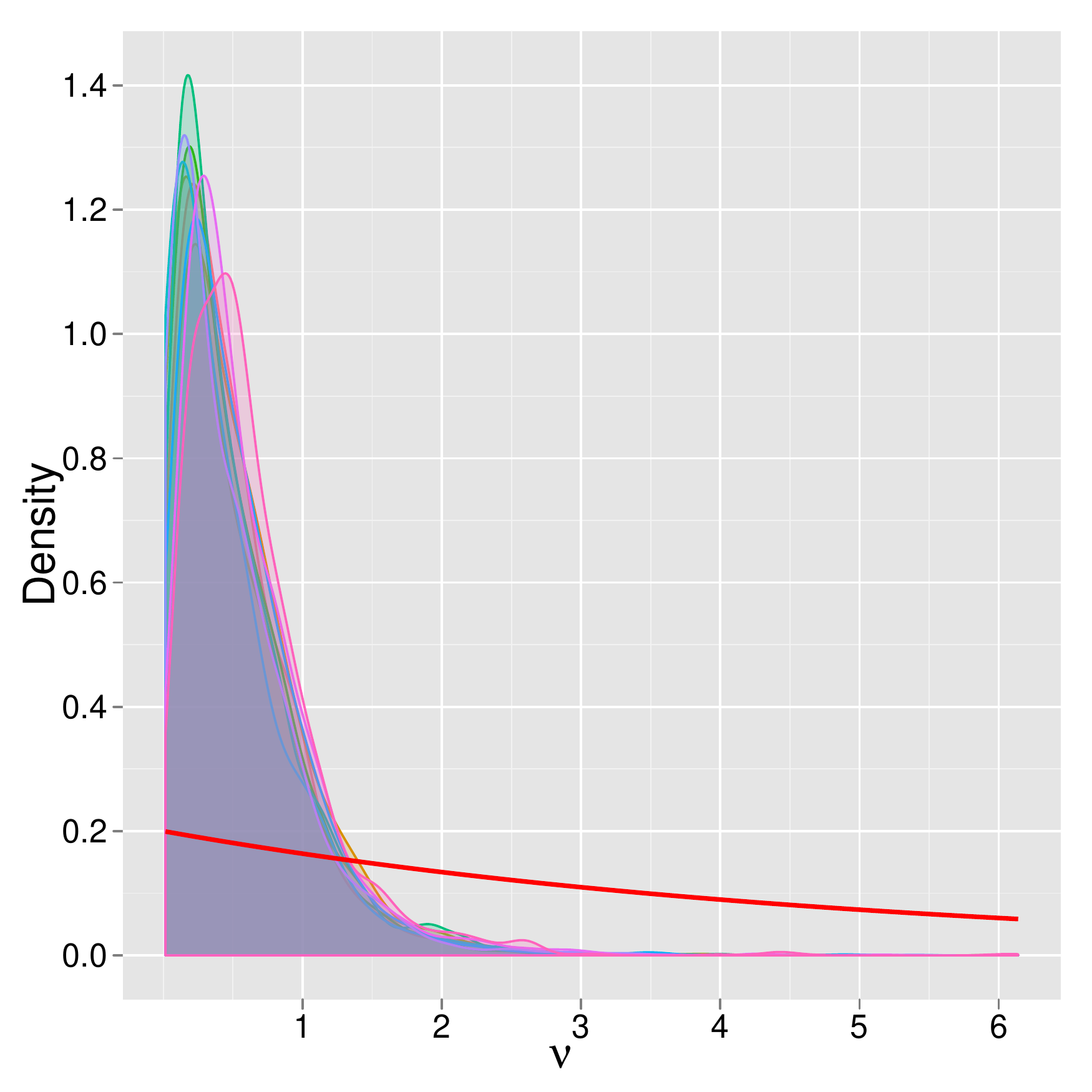}}
 \subfigure[]{\includegraphics[width = 0.3\textwidth]{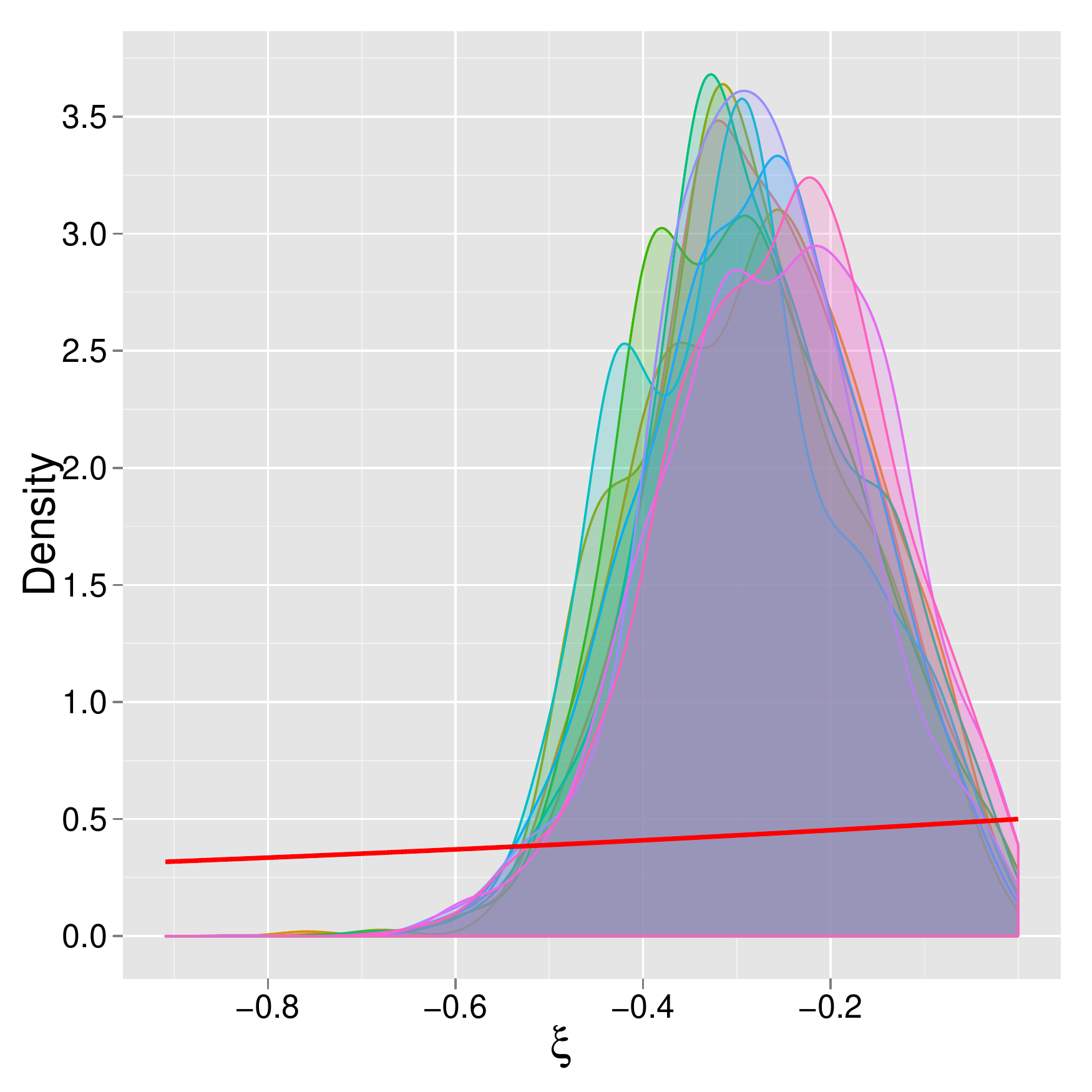}}
 \subfigure[]{\includegraphics[width = 0.3\textwidth]{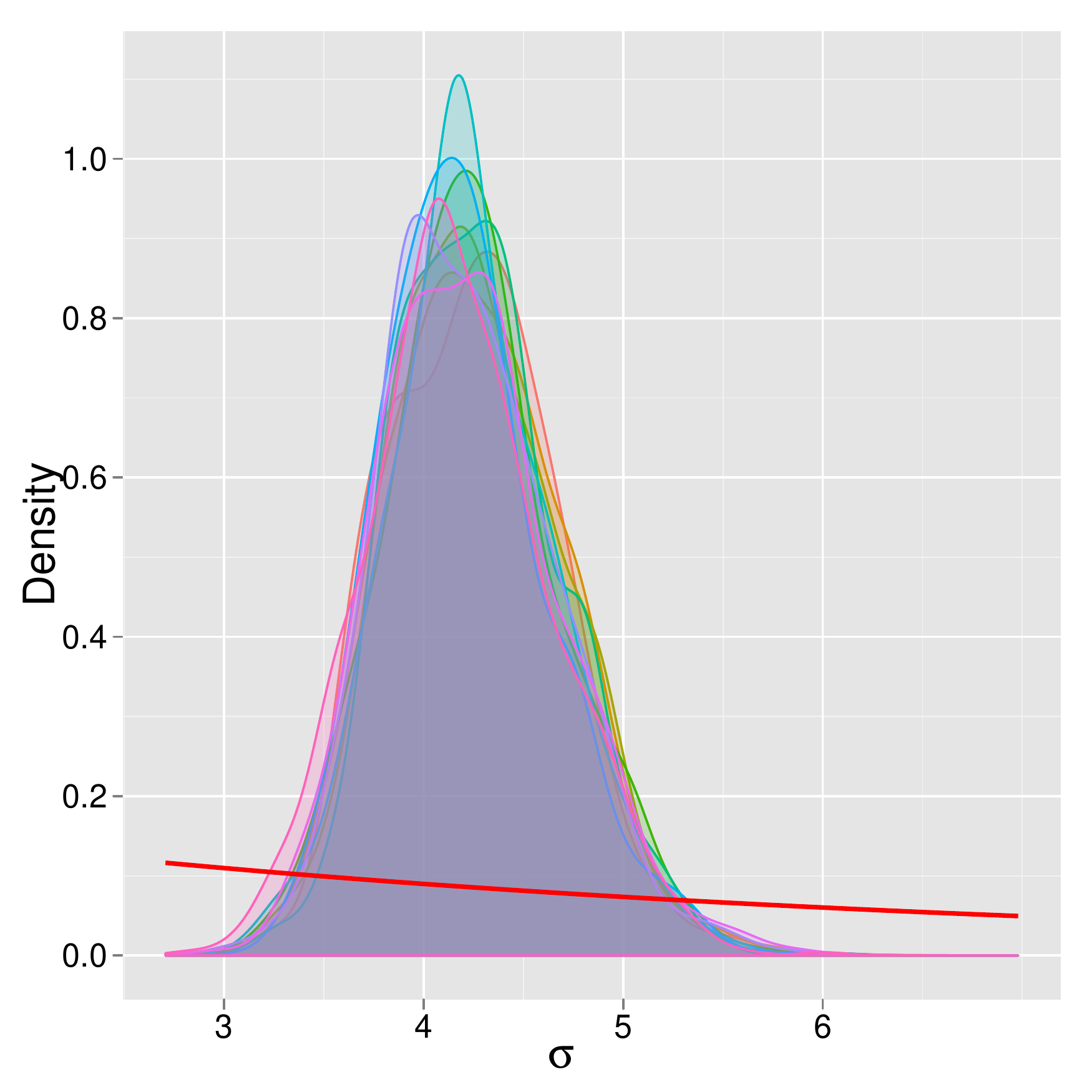}}
 \subfigure[]{\includegraphics[width = 0.3\textwidth]{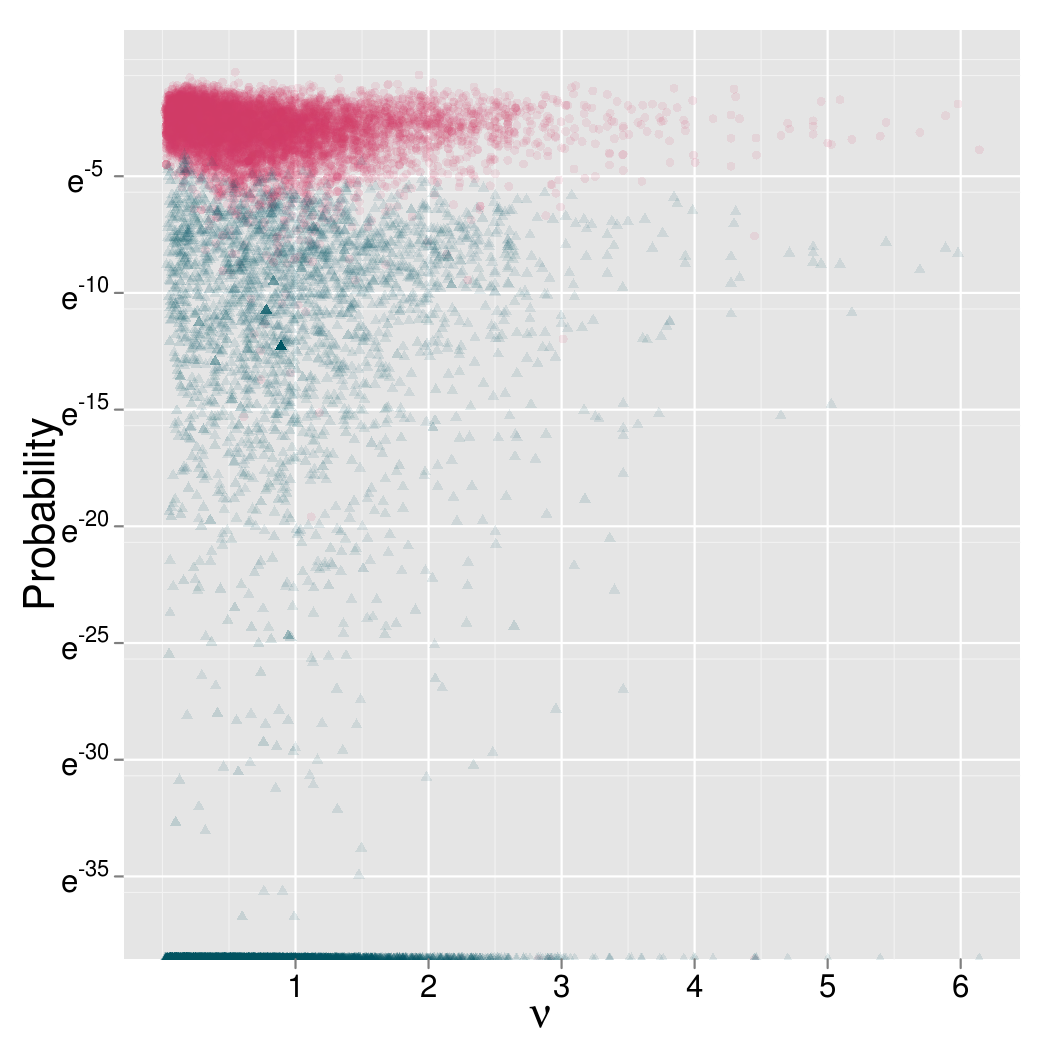}}
 \subfigure[]{\includegraphics[width = 0.3\textwidth]{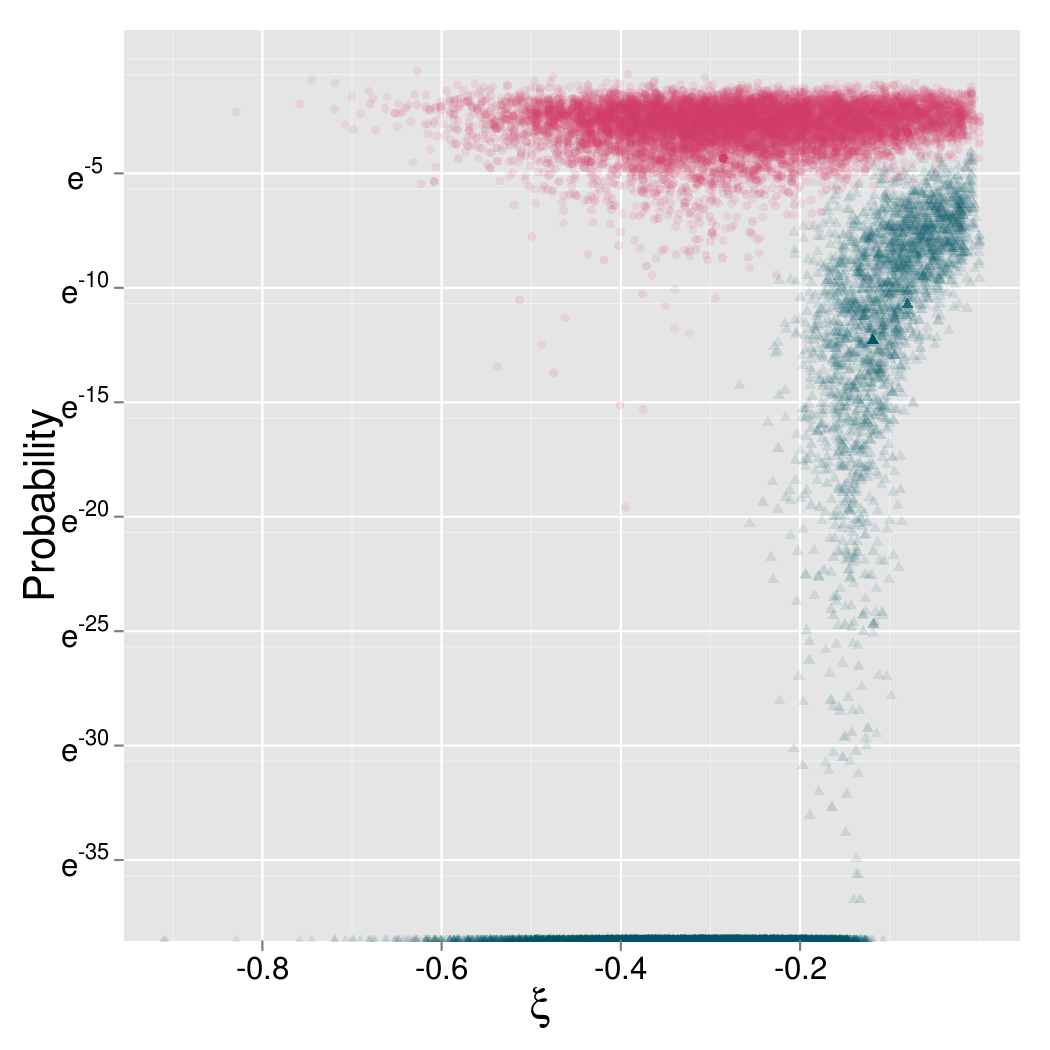}}
 \subfigure[]{\includegraphics[width = 0.3\textwidth]{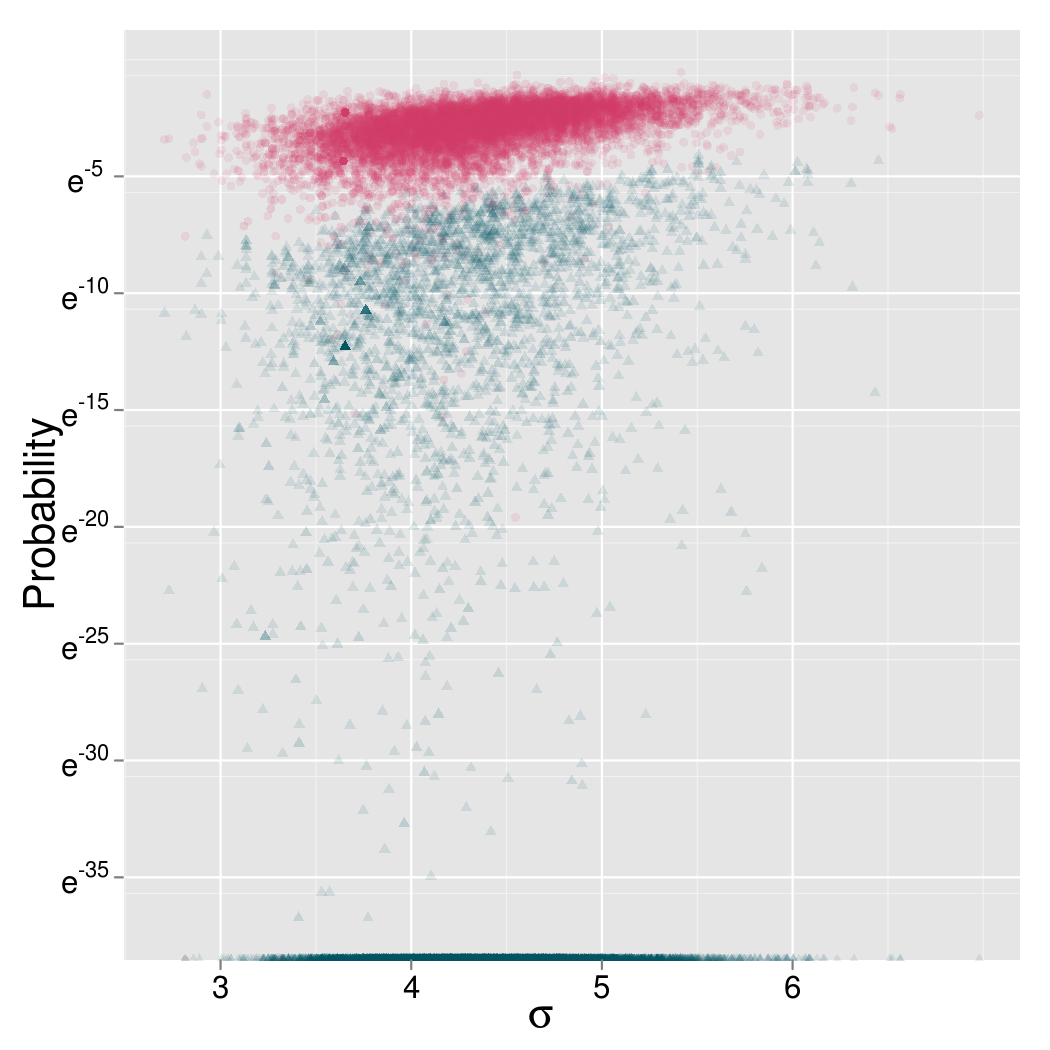}}
 \caption{\label{fig:athletics:perparameter} Athletics records. (a) Best two times of each
year, in women's 3000
metres events between 1976
and 2010; (b) box-plots over 10 runs of \SMCSQ of estimates of the probability of interest
(top) ${p_t^{502.62}}$, (middle) ${p_t^{cond}}$ and
(bottom) ${p_t^{486.11}}$; the
$y$-axis is in log scale and
the dotted line indicates
the year 1993; (c)-(e) posterior distribution of the
parameters approximated by  \SMCSQ, 
with results of $10$ independent runs overlaid on each
plot and where the full line represents the prior density function; 
(f)-(h)  probability of 
observing at year $1993$ a recorded time less than $486.11$ seconds (blue triangles,
lower cloud of points) and
less than $502.62$ seconds (red circles, upper cloud of points) against the
components of $\theta$, where point sizes
and transparencies are proportional to the weights of the $\theta$-particles.}
\end{figure}

The second application illustrates the potential
of \SMCSQ in smoothing while accounting for 
parameter uncertainty. In particular, we consider 
state-space models that have been proposed for the dynamic evolution
of athletic records, see for example
\cite{RobinsonTawn1995},
\cite{GaetanGrigoletto2004}, \cite{fearnhead2010smoothinglinear}. 
We analyse the time series of the best times recorded for women's 3000 metres
running events between 1976 and 2010. The motivation is to assess to which
extent
Wang Junxia's world record in 1993 was unusual: $486.11$ seconds while
the previous  record was $502.62$ seconds. The data  is shown in
Figure \ref{fig:athletics:observations} and  include two observations
per year $y = y_{1:2}$, with $y_1
< y_2$: $y_1$ is the best annual time and $y_2$ the second best time on the race
where $y_1$ was recorded.  The data is available from
\href{http://www.alltime-athletics.com/}{http://www.alltime-athletics.com/}
and it is further discussed in the aforementioned articles. A further fact
that sheds doubt on the record is that the second time for 1993 corresponds to
an athlete from the same team as the record holder.

We use the same modelling as \citet{fearnhead2010smoothinglinear}. The 
observations follow a
generalized extreme value (GEV) distribution for minima, with cumulative
distribution function $G$ defined by:
\begin{equation}
  \label{eq:gev}
  G(y \vert \mu, \xi, \sigma) = 1 - \exp\left[-\left\{1 - \xi
\left(\frac{y - \mu}{\sigma}\right)\right\}_+^{-1/\xi}\right]
\end{equation}
where $\mu$, $\xi$ and $\sigma$ are respectively the location, shape and scale
parameters, and $\{\cdot\}_+ = \max(0, \cdot)$. We denote by $g$ the associated
probability density function. The support of this distribution depends
on the parameters; e.g. if $\xi < 0$, $g$ and $G$ are non-zero
 for $y > \mu + \sigma/\xi$. The probability density function
 for $y=y_{1:2}$ is given by:
 \begin{equation}
   \label{eq:obs-den-ath}
   g(y_{1:2}\vert \mu, \xi, \sigma) = \{1 - G(y_2\vert\mu, \xi, \sigma)\}
\prod_{i=1}^2 \frac{g(y_{i}\vert \mu, \xi, \sigma)}{1 - G(y_{i}\vert \mu,
\xi, \sigma)}\,
 \end{equation}
subject to $y_1<y_2$. 
The location $\mu$ is not treated as a parameter but as a smooth second-order random
walk process:
\begin{equation}
  \label{eq:system-ath}
  x_{t} = (\mu_{t},\dot{\mu}_{t})'\,,\quad x_{t+1} \mid x_t,\nu \sim 
\mathcal{N}\left(F x_{t}, Q \right)\,,\quad 
F = \begin{pmatrix}
  1 & 1  \\
  0 & 1 \end{pmatrix}
\text{ and }
Q = \nu^2 \begin{pmatrix}
  1/3 & 1/2  \\
  1/2 & 1 \end{pmatrix}
\end{equation}
To complete the model specification we set a diffuse initial distribution
$\mathcal{N}(520, 10^2)$
on $\mu_0$. Thus we deal with  bivariate observations in time
$y_t=y_{t,1:2}$, a state-space model with non-Gaussian
observation density given in \eqref{eq:obs-den-ath}, a two-dimensional
state process given in \eqref{eq:system-ath}, and a $3$-dimensional
unknown parameter vector, $\theta =
(\nu, \xi, \sigma)$. We choose independent exponential prior distributions on
 $\nu$ and $\sigma$ with rate
$0.2$. The sign of $\xi$ has determining impact on the support of the
observation density, and the computation of extremal probabilities. For this application, given the form of
\eqref{eq:gev} and the fact that the observations are necessarily bounded from below,  
it makes sense to assume that $\xi\leq 0$, hence we take an exponential prior
distribution on $-\xi$ with rate $0.5$. (We also tried a $N(0,3^2)$ prior, which had some
moderate impact on the estimates presented below, but the corresponding results 
are not reported here.) 

The data we will use in the analysis exclude the two times recorded on
1993. Thus, in an abuse of notation $y_{1976:2010}$ below refers to
the pairs of times for all years but 1993, and in the model we assume
that there was no observation for that year. Formally we want to
estimate probabilities 
\[
 p_t^{y} = \mathbb{P}(y_{t} \leq y | y_{1976:2010}) = 
\int_{\Theta} \int_\mathcal{X} G(y | \mu_{t}, \theta)
p(\mu_t | y_{1976:2010},\theta)
p(\theta | y_{1976:2010})\,
d\mu_{t} d\theta
\]
where the smoothing distribution $p(\mu_{t}\vert y_{1976:2010}, \theta)$ and
the posterior distribution $p(\theta \vert
y_{1976:2010})$ appear explicitly; below we also consider the
probabilities conditionally on the  parameter values, rather than integrating over
those. The interest lies in
$p_{1993}^{486.11}$, $p_{1993}^{502.62}$ and $p_t^{cond} :=
p_t^{486.11} / p_t^{502.62}$, 
which is the probability of observing at year $t$ Wang Junxia's record given
that we observe a better time than the previous world record. 
The rationale
for using this conditional probability is to take into account the exceptional
nature of any new world record.



The algorithm is launched $10$ times with $N_\theta= 1,000$ and $N_x =
250$. The resample-move steps are triggered when the ESS goes below $50\%$,
as in the previous example, and the
proposal distribution used in the move steps is an independent Gaussian
distribution fitted on the particles. The computing time of each of the $10$ runs
varies between $30$ and $70$ seconds (using the same machine as in the previous section), which is why we allowed ourselves to 
use a fairly large number of particles compared to the small time horizon.
Figure \ref{fig:athletics:probabeating} represents the estimates $\hat{p}_t^y$ at each year, for $y =
486.11$ (lower box-plots) and $y = 502.62$ (upper box-plots), as well as
$\hat{p}_t^{cond} = \hat{p}_t^{486.11} / \hat{p}_t^{502.62}$
(middle box-plots). The box-plots show the variability across the independent
runs of the algorithm, and the lines connect the mean values computed across
independent runs at each year. The mean value of
$\hat{p}_{1993}^{cond}$ over the runs is
$9.4\cdot 10^{-4}$ and the standard deviation over the runs is $3.3 \cdot
10^{-4}$. Note that the estimates $\hat{p}_t^y$ are computed using
the smoothing algorithm described in Section \ref{sub:invariant}.  

The second row of Figure
\ref{fig:athletics:perparameter} shows
the posterior distributions of the three parameters $(\nu, \xi, \sigma)$ using
kernel density estimations of the weighted $\theta$-particles. The density
estimators obtained for each
run are overlaid to show the consistency of the results over independent runs.
The prior
density function (full line) is nearly flat over the region of
high posterior mass.
The third row of Figure
\ref{fig:athletics:perparameter} shows scatter plots of the probabilities
$G(y \vert \mu_{1993}^{n^\star(m)}, \theta^m)$ against the parameters
$\theta^m$. The triangles represent these probabilities for $y = 486.11$
while the circles represent the probabilities for $y = 502.62$. The cloud
of points at the bottom of these plots correspond to parameters
$\theta^m$ for which the probability $G(486.11 \vert
\mu_{1993}^{n^\star(m)}, \theta^m)$ is exactly 0.

\section{Extensions} \label{sec:conclusion}

In this paper, we developed an ``exact approximation'' of the
IBIS algorithm, that is, an ideal SMC algorithm targetting the sequence
$\pi_{t}(\theta)=p(\theta|y_{1:t})$, with incremental weight
$\pi_{t}(\theta)/\pi_{t-1}(\theta)=p(y_{t}|y_{1:t-1},\theta)$. The phrase
``exact approximation'', borrowed from \citet{PMCMC}, refers to the fact
that our approach targets the exact marginal distributions, for any fixed
value $N_x$. 

\subsection{Intractable densities}

We have argued that \SMCSQ can cope with state-space models with
intractable transition densities provided these can be simulated
from. More generally, it can cope with intractable transition of
observation densities provided they  can
be unbiasedly estimated. Filtering for dynamic models with intractable densities for which
unbiased estimators can be computed was discussed in
\cite{fear:papa:robe}. It was shown that replacing these densities by
their unbiased estimators is equivalent to introducing additional
auxiliary variables in the state-space model.  \SMCSQ can directly be
applied in this context by replacing these terms by the
unbiased estimators to obtain sequential state and parameter inference
for such models. 

\subsection{\SMCSQ for tempering}

A natural question is whether we can construct other types of \SMCSQ
algorithms, which would be ``exact approximations'' of different
SMC strategies. Consider for instance, again for a state-space model,
the following geometric bridge sequence \citep[in the spirit of
e.g.][]{Neal:AIS},
which allows for a smooth transition from the prior to the posterior:
\[
\pi_{t}(\theta)\propto p(\theta)\left\{ p(y_{1:T}|\theta)\right\}
^{\gamma_{t}},\quad\gamma_{t}=t/L,
\]
where $L$ is the total number of iterations. 
As pointed out by one referee, see also  \cite{FulopDuan},
it is possible to derive some sort of \SMCSQ
algorithm that targets iteratively the sequence 
\[
\pi_{t}(\theta)\propto p(\theta)\left\{ \hat{p}(y_{1:T}|\theta)\right\}
^{\gamma_{t}},\quad\gamma_{t}=t/L,
\]
where $\hat{p}(y_{1:T}|\theta)$ is a particle filtering estimate of the likelihood.
Note that $\left\{ \hat{p}(y_{1:T}|\theta)\right\}
^{\gamma_{t}}$ is not an unbiased estimate of 
 $\left\{ {p}(y_{1:T}|\theta)\right\}
^{\gamma_{t}}$ when $\gamma_t\in(0,1)$. This makes the interpretation of the algorithm 
more difficult, as it cannot be analysed as a noisy, unbiased, version of an ideal algorithm. 
In particular, Proposition \ref{lem:comp} on the complexity of \SMCSQ
cannot be easily extended to the tempering case.
It is also less flexible in terms of PMCMC steps:
for instance, it is not possible to implement the conditional 
SMC step described in Section \ref{sub:condsmc}, or more generally a particle Gibbs step,
because such steps rely on the mixture representation of the target distribution, where
the mixture index is some selected trajectory, 
see \eqref{eq:Andrieuetal_pit}, and this representation does not hold in the tempering case.  
More importantly, this tempering strategy does not make it possible to perform sequential analysis
as the \SMCSQ algorithm discussed in this paper. 

The fact remains that this tempering strategy may prove useful in certain 
non-sequential scenarios, as suggested by the numerical examples of  
 \cite{FulopDuan}. It may be used also for determining MAP (maximum a posteriori) estimators, 
and in particular the maximum likelihood estimator (using a flat prior), 
by letting $\gamma_t\rightarrow +\infty$. 

\section*{Acknowledgements}

N.~Chopin is supported by the ANR grant ANR-008-BLAN-0218 ``BigMC'' of
the French Ministry of research. 
P.E.~Jacob is supported by a PhD fellowship from the AXA Research Fund. 
O.~Papaspiliopoulos would like to acknowledge financial support by the Spanish government
through a ``Ramon y Cajal" fellowship and grant MTM2009-09063. 
The authors are thankful to 
Arnaud Doucet (University of Oxford), 
Peter M\"uller (UT Austin),
Gareth W. Peters (UCL) and the referees
for useful comments.

\bibliographystyle{apalike}
\bibliography{complete}

\begin{thebibliography}{}

\bibitem[Andrieu et~al., 2010]{PMCMC}
Andrieu, C., Doucet, A., and Holenstein, R. (2010).
\newblock Particle {M}arkov chain {M}onte {C}arlo methods.
\newblock {\em J. R. Statist. Soc. B}, 72(3):269--342.

\bibitem[Andrieu and Roberts, 2009]{pseudo}
Andrieu, C. and Roberts, G. (2009).
\newblock {The pseudo-marginal approach for efficient Monte Carlo
  computations}.
\newblock {\em The Annals of Statistics}, 37(2):697--725.

\bibitem[Barndorff-Nielsen and Shephard, 2001]{bns:ou}
Barndorff-Nielsen, O.~E. and Shephard, N. (2001).
\newblock Non-{G}aussian {O}rnstein-{U}hlenbeck-based models and some of their
  uses in financial economics.
\newblock {\em J. R. Stat. Soc. Ser. B Stat. Methodol.}, 63(2):167--241.

\bibitem[Barndorff-Nielsen and Shephard, 2002]{bns:real}
Barndorff-Nielsen, O.~E. and Shephard, N. (2002).
\newblock Econometric analysis of realized volatility and its use in estimating
  stochastic volatility models.
\newblock {\em J. R. Stat. Soc. Ser. B Stat. Methodol.}, 64(2):253--280.

\bibitem[Black, 1976]{black}
Black, F. (1976).
\newblock {Studies of stock price volatility changes}.
\newblock In {\em Proceedings of the 1976 meetings of the business and economic
  statistics section, American Statistical Association}, volume 177, page~81.

\bibitem[Capp\'e et~al., 2004]{Robert:PMC}
Capp\'e, O., Guillin, A., Marin, J.~M., and Robert, C. (2004).
\newblock Population {M}onte {C}arlo.
\newblock {\em J. Comput. Graph. Statist.}, 23:907--929.

\bibitem[Capp\'e et~al., 2005]{CapMouRyd}
Capp\'e, O., Moulines, E., and Ryd\'en, T. (2005).
\newblock {\em Inference in Hidden {M}arkov Models}.
\newblock Springer-Verlag, New York.

\bibitem[Carpenter et~al., 1999]{CarClifFearn}
Carpenter, J., Clifford, P., and Fearnhead, P. (1999).
\newblock Improved particle filter for nonlinear problems.
\newblock {\em IEE Proc. Radar, Sonar Navigation}, 146(1):2--7.

\bibitem[Carvalho et~al., 2010]{carvalho2010particlelearning}
Carvalho, C., Johannes, M., Lopes, H., and Polson, N. (2010).
\newblock {Particle learning and smoothing}.
\newblock {\em Statistical Science}, 25(1):88--106.

\bibitem[C{\'e}rou et~al., 2011]{CerDelGuy2011nonasymptotic}
C{\'e}rou, F., Del~Moral, P., and Guyader, A. (2011).
\newblock A nonasymptotic theorem for unnormalized {F}eynman--{K}ac particle
  models.
\newblock {\em Ann. Inst. Henri Poincarr\'e}, 47(3):629--649.

\bibitem[Chernov et~al., 2003]{chernov:multi}
Chernov, M., Ronald~Gallant, A., Ghysels, E., and Tauchen, G. (2003).
\newblock {Alternative models for stock price dynamics}.
\newblock {\em Journal of Econometrics}, 116(1-2):225--257.

\bibitem[Chopin, 2002]{Chopin:IBIS}
Chopin, N. (2002).
\newblock A sequential particle filter for static models.
\newblock {\em Biometrika}, 89:539--552.

\bibitem[Chopin, 2004]{Chopin:CLT}
Chopin, N. (2004).
\newblock Central {L}imit {T}heorem for sequential {M}onte {C}arlo methods and
  its application to {B}ayesian inference.
\newblock {\em Ann. Stat.}, 32(6):2385--2411.

\bibitem[Chopin, 2007]{Chopin2007}
Chopin, N. (2007).
\newblock Inference and model choice for sequentially ordered hidden {M}arkov
  models.
\newblock {\em J. R. Statist. Soc. B}, 69(2):269--284.

\bibitem[Crisan and Doucet, 2002]{crisan2002survey}
Crisan, D. and Doucet, A. (2002).
\newblock {A survey of convergence results on particle filtering methods for
  practitioners}.
\newblock {\em IEEE J. Sig. Proc.}, 50(3):736--746.

\bibitem[Del~Moral, 2004]{delMoral:book}
Del~Moral, P. (2004).
\newblock {\em Feynman-Kac Formulae}.
\newblock Springer.

\bibitem[Del~Moral et~al., 2006]{DelDouJas:SMC}
Del~Moral, P., Doucet, A., and Jasra, A. (2006).
\newblock {Sequential Monte Carlo samplers}.
\newblock {\em Journal of the Royal Statistical Society: Series B (Statistical
  Methodology)}, 68(3):411--436.

\bibitem[Del~Moral and Guionnet, 1999]{DelGui}
Del~Moral, P. and Guionnet, A. (1999).
\newblock Central limit theorem for nonlinear filtering and interacting
  particle systems.
\newblock {\em Ann. Appl. Prob.}, 9:275--297.

\bibitem[Douc and Moulines, 2008]{douc:moulines:2008}
Douc, R. and Moulines, E. (2008).
\newblock Limit theorems for weighted samples with applications to sequential
  {M}onte {C}arlo methods.
\newblock {\em Ann. Statist.}, 36(5):2344--2376.

\bibitem[Doucet et~al., 2001]{DouFreiGor}
Doucet, A., de~Freitas, N., and Gordon, N.~J. (2001).
\newblock {\em Sequential {M}onte {C}arlo Methods in Practice}.
\newblock Springer-Verlag, New York.

\bibitem[Doucet et~al., 2000]{DouGodAnd}
Doucet, A., Godsill, S., and Andrieu, C. (2000).
\newblock On sequential {M}onte {C}arlo sampling methods for {B}ayesian
  filtering.
\newblock {\em Statist. Comput.}, 10(3):197--208.

\bibitem[Doucet et~al., 2009]{Doucetetal2009OverviewParameterEstimation}
Doucet, A., Kantas, N., Singh, S., and Maciejowski, J. (2009).
\newblock An overview of {S}equential {M}onte {C}arlo methods for parameter
  estimation in general state-space models.
\newblock In {\em Proceedings IFAC System Identification (SySid) Meeting.}

\bibitem[Doucet et~al., 2011]{DoucPoyiaSin}
Doucet, A., Poyiadjis, G., and Singh, S. (2011).
\newblock Sequential {M}onte {C}arlo computation of the score and observed
  information matrix in state-space models with application to parameter
  estimation.
\newblock {\em Biometrika}, 98:65--80.

\bibitem[Fearnhead, 2002]{Fearnhead:Sufficient}
Fearnhead, P. (2002).
\newblock {MCMC}, sufficient statistics and particle filters.
\newblock {\em Statist. Comput.}, 11:848--862.

\bibitem[Fearnhead et~al., 2010a]{high}
Fearnhead, P., Papaspiliopoulos, O., Roberts, G., and Stuart, A. (2010a).
\newblock Random weight particle filtering of continuous time processes.
\newblock {\em J. R. Stat. Soc. Ser. B Stat. Methodol.}, 72:497--513.

\bibitem[Fearnhead et~al., 2008]{fear:papa:robe}
Fearnhead, P., Papaspiliopoulos, O., and Roberts, G.~O. (2008).
\newblock Particle filters for partially observed diffusions.
\newblock {\em J. R. Statist. Soc. B}, 70:755--777.

\bibitem[Fearnhead et~al., 2010b]{fearnhead2010smoothinglinear}
Fearnhead, P., Wyncoll, D., and Tawn, J. (2010b).
\newblock {A sequential smoothing algorithm with linear computational cost}.
\newblock {\em Biometrika}, 97(2):447.

\bibitem[Fulop and Duan, 2011]{FulopDuan}
Fulop, A. and Duan, J. (2011).
\newblock Marginalized sequential monte carlo samplers.
\newblock Technical report, SSRN 1837772.

\bibitem[Fulop and Li, 2011]{FulopLi}
Fulop, A. and Li, J. (2011).
\newblock Robust and efficient learning: A marginalized resample-move approach.
\newblock Technical report, SSRN 1724203.

\bibitem[Gaetan and Grigoletto, 2004]{GaetanGrigoletto2004}
Gaetan, C. and Grigoletto, M. (2004).
\newblock {Smoothing sample extremes with dynamic models}.
\newblock {\em Extremes}, 7(3):221--236.

\bibitem[Gilks and Berzuini, 2001]{GilksBerzu}
Gilks, W.~R. and Berzuini, C. (2001).
\newblock Following a moving target - {M}onte {C}arlo inference for dynamic
  {B}ayesian models.
\newblock {\em J. R. Statist. Soc. B}, 63:127--146.

\bibitem[Gordon et~al., 1993]{Gordon}
Gordon, N.~J., Salmond, D.~J., and Smith, A. F.~M. (1993).
\newblock Novel approach to nonlinear/non-{G}aussian {B}ayesian state
  estimation.
\newblock {\em IEE Proc. F, Comm., Radar, Signal Proc.}, 140(2):107--113.

\bibitem[Griffin and Steel, 2006]{grif:steel:ou}
Griffin, J. and Steel, M. (2006).
\newblock {Inference with non-Gaussian Ornstein-Uhlenbeck processes for
  stochastic volatility}.
\newblock {\em Journal of Econometrics}, 134(2):605--644.

\bibitem[Harvey and Shephard, 1996]{harvey:shep}
Harvey, A. and Shephard, N. (1996).
\newblock {Estimation of an asymmetric stochastic volatility model for asset
  returns}.
\newblock {\em Journal of Business \& Economic Statistics}, 14(4):429--434.

\bibitem[Jasra et~al., 2007]{jasra2007population}
Jasra, A., Stephens, D., and Holmes, C. (2007).
\newblock {On population-based simulation for static inference}.
\newblock {\em Statistics and Computing}, 17(3):263--279.

\bibitem[Kim et~al., 1998]{kim1998stochastic}
Kim, S., Shephard, N., and Chib, S. (1998).
\newblock {Stochastic volatility: likelihood inference and comparison with ARCH
  models}.
\newblock {\em Rev. Econ. Studies}, 65(3):361--393.

\bibitem[Kitagawa, 1998]{Kitagawa:Self}
Kitagawa, G. (1998).
\newblock A self-organizing state-space model.
\newblock {\em J. Am. Statist. Assoc.}, 93:1203--1215.

\bibitem[Koop and Potter, 2007]{Koop2007}
Koop, G. and Potter, S.~M. (2007).
\newblock Forecasting and estimating multiple change-point models with an
  unknown number of change-points.
\newblock {\em Review of Economic Studies}, 74:763 -- 789.

\bibitem[K{\"u}nsch, 2001]{Kun:SSHMM}
K{\"u}nsch, H. (2001).
\newblock State space and hidden {M}arkov models.
\newblock In Barndorff-Nielsen, O.~E., Cox, D.~R., and Kl{\"u}ppelberg, C.,
  editors, {\em Complex Stochastic Systems}, pages 109--173. Chapman and Hall.

\bibitem[Liu and Chen, 1998]{LiuChen}
Liu, J. and Chen, R. (1998).
\newblock Sequential {M}onte {C}arlo methods for dynamic systems.
\newblock {\em J. Am. Statist. Assoc.}, 93:1032--1044.

\bibitem[Liu and West, 2001]{LiuWest}
Liu, J. and West, M. (2001).
\newblock Combined parameter and state estimation in simulation-based
  filtering.
\newblock In Doucet, A., de~Freitas, N., and Gordon, N.~J., editors, {\em
  Sequential {M}onte {C}arlo Methods in Practice}, pages 197--223.
  Springer-Verlag.

\bibitem[Liu, 2008]{liu:book}
Liu, J.~S. (2008).
\newblock {\em Monte {C}arlo strategies in scientific computing}.
\newblock Springer Series in Statistics. Springer, New York.

\bibitem[Neal, 2001]{Neal:AIS}
Neal, R.~M. (2001).
\newblock Annealed importance sampling.
\newblock {\em Statist. Comput.}, 11:125--139.

\bibitem[Oudjane and Rubenthaler, 2005]{OudRub}
Oudjane, N. and Rubenthaler, S. (2005).
\newblock Stability and uniform particle approximation of nonlinear filters in
  case of non ergodic signals.
\newblock {\em Stochastic Analysis and applications}, 23:421--448.

\bibitem[Papaspiliopoulos et~al., 2007]{pap:rob:sk}
Papaspiliopoulos, O., Roberts, G.~O., and Sk{\"o}ld, M. (2007).
\newblock A general framework for the parametrization of hierarchical models.
\newblock {\em Statist. Sci.}, 22(1):59--73.

\bibitem[Peters et~al., 2010]{PetersHosackHayes}
Peters, G., Hosack, G., and Hayes, K. (2010).
\newblock Ecological non-linear state space model selection via adaptive
  particle markov chain monte carlo.
\newblock {\em Arxiv preprint arXiv:1005.2238}.

\bibitem[Roberts et~al., 2004]{robe:papa:dell:2004}
Roberts, G.~O., Papaspiliopoulos, O., and Dellaportas, P. (2004).
\newblock Bayesian inference for non-{G}aussian {O}rnstein-{U}hlenbeck
  stochastic volatility processes.
\newblock {\em J. R. Stat. Soc. Ser. B Stat. Methodol.}, 66(2):369--393.

\bibitem[Robinson and Tawn, 1995]{RobinsonTawn1995}
Robinson, M. and Tawn, J. (1995).
\newblock {Statistics for exceptional athletics records}.
\newblock {\em Applied Statistics}, 44(4):499--511.

\bibitem[Silva et~al., 2009]{silva2009particle}
Silva, R., Giordani, P., Kohn, R., and Pitt, M. (2009).
\newblock Particle filtering within adaptive {M}etropolis {H}astings sampling.
\newblock {\em Arxiv preprint arXiv:0911.0230}.

\bibitem[Storvik, 2002]{Storvik}
Storvik, G. (2002).
\newblock Particle filters for state-space models with the presence of unknown
  static parameters.
\newblock {\em IEEE Transaction on Signal Processing}, 50:281--289.

\bibitem[Whiteley, 2011]{whiteley2011stability}
Whiteley, N. (2011).
\newblock Stability properties of some particle filters.
\newblock {\em Arxiv preprint arXiv:1109.6779}.

\end{thebibliography}

\section*{Appendix A: Proof of Proposition \eqref{lem:pit}}

We remark first that $\psi_{t,\theta}$ may be rewritten as follows: 
 \[
\psi_{t,\theta}(x_{1:t}^{1:N_x},a_{1:t-1}^{1:N_x})=\left\{
\prod_{n=1}^{N_x}q_{1,\theta}\left(x_{1}^{n}\right)\right\} \left\{
\prod_{s=2}^{t}\prod_{n=1}^{N_x}W_{s-1,\theta}^{a_{s-1}^{n}}q_{s,\theta}\left(x_{s}^{n}|x_{
s-1}^{a_{s-1}^{n} }\right)\right\} .\]
Starting from \eqref{eq:def_pit} and \eqref{eq:Zhatt}, one obtains
\begin{eqnarray}
\pi_{t}(\theta,x_{1:t}^{1:N_x},a_{1:t-1}^{1:N_x}) 
& = &
\frac{p(\theta)\psi_{t,\theta}(x_{1:t}^{1:N_x},a_{1:t-1}^{1:N_x})}{N_x^t p(y_{1:t})}
\prod_{s=1}^{t}\left\{\sum_{n=1}^{N_x}w_{s,\theta}(x_{s-1}^{a_{s-1}^n},x_s^n)\right\}   \nonumber\\
& = & \frac{p(\theta)}{N_x^t p(y_{1:t})}\sum_{n=1}^{N_x}\Bigg[w_{t,\theta}(x_{t-1}^{a_{t-1}^n},x_t^n )\left\{
\prod_{i=1}^{N_x}q_{1,\theta}(x_{1}^{i})\right\} \nonumber \\
 &  & \qquad\left\{
\prod_{s=2}^{t}\prod_{i=1}^{N_x}W_{s-1,\theta}^{a_{s-1}^i}q_{s,\theta}(x_s^i|x_{s-1}^
{a_{s-1}^i})\right\} 
\prod_{s=1}^{t-1}\left\{
\sum_{i=1}^{N_x}w_{s,\theta}(x_{s-1}^{a_{s-1}^i},x_{s}^i)\right\}
\Bigg] 
\nonumber
\end{eqnarray}
by distributing the final product in the first line, and using 
the convention that $w_{1,\theta}(x_0^{a_0^n}, x_1^n) = w_{1,\theta}( x_1^n) $. 

To obtain \eqref{eq:prop_pit}, we consider the summand
above, for a given value of $n$,  and put aside the random variables that correspond to
the state trajectory $\x_{1:t}^{n}$. We start with
 $\x_{1:t}^{n}(t)=x_{t}^{n}$,
and note that 
\begin{eqnarray*}
w_{t,\theta}(x_{t-1}^{a_{t-1}^n},x_t^n)
q_{t,\theta}(x_{t}^{n}|x_{t-1}^{a_{t-1}^{n}}) & = &
f_{\theta}(x_{t}^{n}|x_{t-1}^{a_{t-1}^{n}})
g_{\theta}(y_{t}|x_{t}^{n})\\
 & = &
f_{\theta}\left(\x_{1:t}^{n}(t)|\x_{1:t}^{n}(t-1)\right)
g_{\theta}\left(y_{t} |\x_{1:t}^{n}(t)\right),
\end{eqnarray*}
and that 
$$
W_{t-1,\theta}^{a_{t-1}^n} 
\left\{ \sum_{i=1}^{N_x} w_{t-1,\theta}(x_{t-2}^{a_{t-2}^i},x_{t-1}^i) \right\}
= w_{t-1,\theta}\left( \x_{1:t}^n(t-2), \x_{1:t}^n(t-1) \right).  
$$
Thus, the summand in the expression of $\pi_t$ above may be rewritten as 
\begin{multline*}
w_{t-1,\theta}(\x_{1:t}^n(t-2),\x_{1:t}^n(t-1))
f_{\theta}\left(\x_{1:t}^{n}(t)|\x_{1:t}^{n}(t-1)\right)g_{\theta}\left(y_{t}
|\x_{1:t}^{n}(t)\right)
\left\{ \prod_{i=1}^{N_x}q_{1,\theta}(x_{1}^{i})\right\}
\times
\\ 
\left\{
\prod_{s=2}^{t-1}\prod_{i=1}^{N_x}W_{s-1,\theta}^{a_{s-1}^i}q_{s,\theta}(x_s^i|x_{s-1}^
{a_{s-1}^i})\right\} 
\left\{\prod_{\stackrel{i=1}{i\neq \h_t^n(n)} }^{N_x}
W_{t-1,\theta}^{a_{t-1}^i}q_{t,\theta}(x_t^i|x_{t-1}^{a_{t-1}^i}) \right\}
\prod_{s=2}^{t-1}\left\{\sum_{i=1}^{N_x}w_{s-1,\theta}(x_{s-2}^{a_{s-2}^n},x_{s-1}^i)\right\}.
\end{multline*}

By applying recursively, for $s=t-1,\ldots,1$ the same type of substitutions,
that is, 
\begin{eqnarray*}
w_s \left(\x_{1:t}^n(s-1),\x_{1:t}^n(s) \right) 
q_{s,\theta} ( x_s^{\h_t^n(s)} | x_{s-1}^{\h_t^n(s-1)} ) & = &
f_{\theta}\left(\x_{1:t}^{n}(s)|\x_{1:t}^{n}(s-1)\right)
g_{\theta}\left(y_s |\x_{1:t}^{n}(s)\right), \\
w_1 \left(\x_{1}^n(1) \right) 
q_{1,\theta} ( x_1^{\h_t^n(1)} ) & = &
\mu_{\theta}\left(\x_{1:t}^{n}(1)\right)
g_{\theta}\left(y_1 |\x_{1:t}^{n}(1)\right), 
\end{eqnarray*}
and, for $s\geq 2 $, 
$$
W_{s-1,\theta}^{a_{s-1}^n} 
\left\{ \sum_{i=1}^{N_x} w_{s-1,\theta}(x_{s-2}^{a_{s-2}^i},x_{s-1}^i) \right\}
= w_{s-1,\theta}\left( \x_{1:t}^n(s-2), \x_{1:t}^n(s-1) \right).  
$$
and noting that 
\begin{eqnarray*}
p(\theta,\x_{1:t}^n,y_{1:t})
& = & p(y_{1:t}) p(\theta|y_{1:t}) p(\x_{1:t}^n|y_{1:t},\theta) \\
& = & p(\theta) \prod_{s=1}^t \left\{
f_\theta \left(\x_{1:t}^{n}(s)|\x_{1:t}^{n}(s-1)\right)
g_{\theta}\left(y_s |\x_{1:t}^{n}(s)\right) 
\right\},
\end{eqnarray*}
where $p(\theta,\x_{1:t}^{n},y_{1:t})$ stands for the joint probability
density defined by the model, for the triplet of random variables
$(\theta,x_{1:t},y_{1:t})$, evaluated at $x_{1:t}=\x_{1:t}^{n}$, 
one eventually gets: 
\begin{eqnarray*}
\lefteqn{\pi_{t}(\theta,x_{1:t}^{1:N_x},a_{1:t-1}^{1:N_x})=p(\theta|y_{1:t})
\times}\\
 &  & \frac{1}{N_x^t} \sum_{n=1}^{N_x}p(\x_{1:t}^{n}|\theta,y_{1:t})\left\{ \myprodone
q_{1,\theta}(x_{1}^{i})\right\} \left\{ \prod_{s=2}^{t}\myprod
W_{s-1,\theta}^{a_{s-1}^{i}}q_{s,\theta}(x_{s}^{i}|x_{s-1}^{a_{s-1}^{i}})\right\}.
\end{eqnarray*}

\section*{Appendix B: Proof of Proposition \ref{lem:comp} and
  discussion of assumptions}

Since  $p(\theta|y_{1:t})$ is the marginal
distribution of $\bar{\pi}_{t,t+p}$, by iterated conditional expectation we get: 
\begin{eqnarray*}
\frac{p(y_{t+1:t+p}|y_{1:t})^{2}}{\mathcal{E}_{t,t+p}^{N_{x}}}
 & = & \mathbb{E}_{p(\theta|y_{1:t})}\left[\mathbb{E}_{\bar{\pi}_{t,t+p}(\cdot|\theta)}\left\{ \hat{Z}_{t+p|t}^{2}(\theta,x_{1:t+p}^{1:N_{x}},a_{1:t+p-1}^{1:N_{x}})\right\} \right]\\
 & = & \mathbb{E}_{p(\theta|y_{1:t})}\left[\mathbb{E}_{\pi_{t}(\cdot|\theta)}\left\{ \mathbb{E}_{\psi_{t+p|t}(\cdot|x_{1:t}^{1:N_{x}},a_{1:t-1}^{1:N_{x}}\theta)}\left\{ \hat{Z}_{t+p|t}^{2}(\theta,x_{1:t+p}^{1:N_{x}},a_{1:t+p-1}^{1:N_{x}})\right\} \right\} \right]\,.
\end{eqnarray*}

To study the inner expectation, we make the following first set of
assumptions: 

\begin{itemize}
 \item[(H1a)]  For all $\theta\in\Theta$, and $x,x',x''\in\mathcal{X}$, 
\[
\frac{f_{\theta}(x|x')}{f_{\theta}(x|x'')}\leq\beta.
\]

\item[(H1b)] For all $\theta\in\Theta$, $x,x'\in\mathcal{X}$, $y\in\mathcal{Y}$,
\[
\frac{g_{\theta}(y|x)}{g_{\theta}(y|x')}\leq\delta.
\]
\end{itemize}

Under these assumptions,  one  obtains  the following non-asymptotic bound.

\begin{lem}[Theorem 1.5 of  \cite{CerDelGuy2011nonasymptotic}]
\label{non-as}
For $N_{x}\geq\beta\delta p$, 
\[
\mathbb{E}_{\psi_{t+p|t}(\cdot|x_{1:t}^{1:N_{x}},a_{1:t-1}^{1:N_{x}}\theta)}
\left\{ \frac{\hat{Z}_{t+p|t}^{2}(\theta,x_{1:t+p}^{1:N_{x}},a_{1:t+p-1}^{1:N_{x}})}{p(y_{t+1:t+p}|y_{1:t},\theta)^{2}}\right\} -1
\leq 4\beta\delta\frac{p}{N_{x}}.
\]
\end{lem}

The Proposition above is taken from \cite{CerDelGuy2011nonasymptotic}, up to some change of notations
and a minor modification: \cite{CerDelGuy2011nonasymptotic} establish this result for the likelihood
estimate $\hat{Z}_{t}$, obtained by running a particle from time
$1$ to time $t$. However, their proof applies straightforwardly
to the partial likelihood estimate $\hat{Z}_{t+p|t}$, obtained by
running a particle filter from time $t+1$ to time $t+p$, and therefore
with initial distribution $\eta_{0}$ set to the mixture of Dirac
masses at the particle locations at time $t$. We note in passing
that Assumptions (H1a) and (H1b) may be loosened up slightly, see
\cite{whiteley2011stability}.   
A direct consequence of  Proposition \ref{non-as}, the main
definitions and the iterated expectation is Proposition 2(a) for
$\eta=4 \beta \delta$. 

Proposition 2(b) requires a  second set of conditions 
taken from \cite{Chopin:IBIS}. These  relate to the asymptotic behaviour
of the marginal posterior distribution $p(\theta|y_{1:t})$ and they
have  been used to study the weight degeneracy of IBIS.  Let
$l_{t}(\theta)=\log p(y_{1:t}|\theta)$. The following assumptions
hold almost surely.

\begin{itemize}
\item[(H2a)] The MLE $\hat{\theta}_{t}$ (the mode of function $l_{t}(\theta)$)
exists and converges to $\theta_{0}$ as $n\rightarrow+\infty$.

\item[(H2b)] The observed information matrix defined as 
\[
\Sigma_{t}=-\frac{1}{t}\frac{\partial l_{t}(\hat{\theta}_{t})}{\partial\theta\partial\theta'}
\]
is positive definite and converges to $I(\theta_{0})$, the Fisher
information matrix. 

\item[(H2c)] There exists $\Delta$ such that, for $\delta\in(0,\Delta)$,
\[
\limsup_{t\rightarrow+\infty}\left[\frac{1}{t}
\sup_{\left\Vert \theta-\hat{\theta}_{t}\right\Vert >\delta}\left\{ l_{t}(\theta)-l_{t}(\hat{\theta}_{t})\right\} \right]<0.
\]

\item[(H2d)] The function $l_{t}/t$ is six-times continuously differentiable,
and its derivatives of order six are bounded relative to $t$ over
any compact set $\Theta'\subset\Theta$. 
\end{itemize}

Under these conditions one may apply Theorem 1 of \cite{Chopin:IBIS} 
 \citep[see also Proof of Theorem 4 in][]{Chopin:CLT} 
to conclude that $\mathcal{E}_{t,t+p}^{\infty}\geq 2\gamma $ for a given $\gamma>0$ and 
$t$ large enough, provided $p=\left\lceil \tau t\right\rceil$
for some $\tau>0$ (that depends on $\gamma$).  
Together with Proposition 2(a) and by a small modification of the Proof of Theorem 4 in
\cite{Chopin:CLT} to fix $\gamma$ instead of $\tau$, we obtain Proposition
2(b) provided $N_{x}=\lceil\eta t\rceil$, and
$\eta=4\beta\delta$.

Note that (H2a) and (H2b) essentially amount to establishing that
the MLE has a standard asymptotic behaviour (such as in the IID case).
This type of results for state-space models is far from trivial, owning
among other things to the intractable nature of the likelihood $p(y_{1:t}|\theta)$.
A good entry in this field is Chapter 12 of \cite{CapMouRyd}, where
it can be seen that the first set of conditions above, (H1a) and (H2b),
are sufficient conditions for establishing (H2a) and (H2b), see in
particular Theorem 12.5.7 page 465. Condition (H2d) is trivial to
establish, if one assumes bounds similar to those in (H1a) and (H1b)
for the derivatives of $g_{\theta}$ and $f_{\theta}$. Condition
(H2c) is harder to establish. We managed to prove that this condition
holds for a very simple linear Gaussian model; notes are available
from the first author. Independent work by Judith Rousseau and Elisabeth
Gassiat is currently carried out on the asymptotic properties of posterior
distributions of state-space models, where (H2c) is established under
general conditions (personal communication). 

The implication of Proposition 2 to the stability of \SMCSQ is based
on the additional assumption  that after resampling at time $t$ we obtain exact
samples from $\pi_t$. In practice,  this is only approximately
true since an MCMC scheme is used to sample new particles. This
assumption also underlies the analysis of IBIS in
\cite{Chopin:IBIS}, where it was demonstrated empirically (see
e.g. Fig. 1(a) in that paper) that the MCMC kernel which updates the
$\theta$ particles has a stable efficiency over time since it uses the population of
$\theta$-particles to design the proposal distribution. We also
observe empirically that the performance of the PMCMC step does not
deteriorate over time provided $N_x$ is increased appropriately, see
for example Figure \ref{fig:SVonefactor:Concentration}(b). It is  important to
establish such a result theoretically, i.e., that the total variation
distance of the PMCMC kernel from the target distribution remains bounded
over time provided $N_x$ is increased appropriately. Note that a  fundamental
difference between IBIS and \SMCSQ is that respect is that in the latter the MCMC step targets
distributions in increasing dimensions as time increases. Obtaining
such a theoretical result is a research project on its own right,
since such quantitative results lack, to the best of our knowledge, 
from  the existing literature. The closest in spirit is Theorem 6 in
\cite{pseudo} which, however, holds for ``large enough'' $N_x$,
instead of providing a quantification of how large $N_x$ needs to be.


\end{document}